\documentclass[times,twocolumn,final,authoryear]{elsarticle}
\usepackage{jasr}
\usepackage{framed,multirow}

\usepackage{amssymb}
\usepackage{latexsym}
\usepackage{afterpage}
\usepackage{lscape}
\DeclareUnicodeCharacter{0308}{}

\usepackage{url}
\usepackage{xcolor}
\definecolor{newcolor}{rgb}{.8,.349,.1}

\usepackage[citebordercolor=white]{hyperref}
\usepackage{footmisc}
\usepackage{pdflscape}
\journal{Advances in Space Research}

\begin{document}

\verso{Bhuvana \textit{et al}}

\begin{frontmatter}

\title{Multi-mission view of extragalactic black hole X-ray binaries LMC X-1 and LMC X-3: evolution of broadband spectral features}%
%\tnotetext[tnote1]{This is an example for title footnote coding.}

\author[1]{Bhuvana \snm{G. R.}\corref{cor1}}
\ead{bhuvanahebbar@gmail.com}
%\cortext[cor1]{Corresponding author: 
  %Tel.: +0-000-000-0000;  
  %fax: +0-000-000-0000;}
\author[1]{Radhika \snm{D.}}
\ead{radhikad.isac@gmail.com}
%\fntext[fn1]{This is author footnote for second author.}
\author[2]{Anuj \snm{Nandi}}
%% Third author's email
\ead{anuj@ursc.gov.in}

\address[1]{Department of Physics, Dayananda Sagar University, Hosur Main Road, Bengaluru, 560068, India}
\address[2]{Space Astronomy Group, ISITE Campus, U. R. Rao Satellite Center, Outer Ring Road, Marathahalli, Bengaluru, 560037, India.}

\received{}
\finalform{}
\accepted{}
\availableonline{}
\communicated{}

\begin{abstract}
%%%
Extragalactic black hole X-ray binaries LMC X-1 and LMC X-3 are the persistent sources which are usually found in a soft spectral state. In this study, we consider multi-mission (\textit{MAXI}, \textit{NICER}, \textit{NuSTAR} and \textit{AstroSat}) X-ray observations of both these sources carried out during 2014 $-$ 2020 to perform a detailed spectral and temporal analysis. Study of long term \textit{MAXI} lightcurve shows that the flux variability of LMC X-1 is moderate (fractional variability, $F_{var}\sim20\%$) whereas that of LMC X-3 is high ($F_{var}\sim50\%$) in 2$-$10 keV which is associated with the change in spectral state. The energy spectra of LMC X-1 and LMC X-3 are characterized by a multi-color disc blackbody and a Comptonization component, with LMC X-1 having an additional Fe-line emission feature. \textit{NICER} ($0.3-10$ keV), \textit{NuSTAR} ($3.0-40$ keV) and \textit{AstroSat} ($0.5-20$ keV) spectral analysis collectively shows that LMC X-1 has remained in the soft state (disc flux contribution in 0.3 $-$ 40 keV, $f_{disc}>80\%$, photon index $\Gamma\sim2.06-4.08$) throughout the period of 2014 $-$ 2020. Mass accretion rate ($\dot{M}$) of LMC X-1 during this period, calculated from bolometric luminosity ($0.1 - 50$ keV) is found to be within $0.07-0.24$ $\dot{M}_{Edd}$ (Eddington mass accretion rate). Although LMC X-3 remained in the soft state during most of the observations ($f_{disc}>95\%$, $\Gamma\sim2.3$), it exhibits a transition into hard state ($f_{disc}\sim26\%,\Gamma\sim1.6$) and intermediate state ($f_{disc}=47-73\%,\Gamma\sim2.02-2.36$). $\dot{M}$ of LMC X-3 through different spectral states varies in the range $0.01-0.42$ $\dot{M}_{Edd}$. Temporal studies carried out for these observations show that the Power Density Spectra (PDS) in $0.3-10$ keV follow a red-noise with fractional rms of $\sim2\%$ for LMC X-1 and in case of LMC X-3, rms is less during the soft state ($\sim0.08-2.35\%$), but relatively high in the intermediate ($\sim3.05-4.91\%$) and hard states ($\sim17.06\%$). From spectral modeling of the soft X-ray continuum of \textit{NICER} and \textit{NuSTAR} energy spectra with relativistic accretion disc and reflection models, we constrain the spin and accretion rate ($\dot{M}$) of the BHs. In case of LMC X-1, spin is estimated to be within $0.85-0.94$ by continuum-fitting method and $0.93-0.94$ by Fe-line fitting method and in case of LMC X-3 continuum-fitting yields its value is  in the range $0.16-0.33$. Finally, we discuss the implications of our findings in the context of accretion disc dynamics around the vicinity of the BHs.
%%%%
\end{abstract}


\end{frontmatter}

%% For linenumbers
%\linenumbers

\section{Introduction}
\label{sec1}
The stellar binary systems consisting of a black hole as an accretor and a normal star as its companion are known as Black Hole X-ray Binaries (BH-XRBs). Accretion from the companion star onto the black hole results in the formation of a disc around it which is bright in X-rays. Majority of the BH-XRBs spend most of their time in quiescent state where the X-ray luminosity from the source is very low; beyond detection limit ($< 10^{35}$ erg cm$^{-2}$ sec$^{-1}$) \citep{2006csxs.book..157M}. Such sources are detected in X-rays only when they undergo an outburst where excess flux is emitted in X-rays and hence they are well known as transient X-ray binaries. There are a few XRBs which are always luminous in X-rays and does not attain quiescence and are termed as persistent sources \citep{1997ApJ...491..312C,2016ApJS..222...15T,2016A&A...587A..61C,2018JApA...39....5S}.  \par
While the soft X-ray continuum of the energy spectrum originates from the thermally dominated accretion disc, hard X-rays are produced due to inverse Comptonization of soft photons by the hot corona. Depending on which of these components are dominant in the energy spectra, various spectral states of the BH-XRBs are defined \citep{1995xrbi.nasa..126T,1995ApJ...455..623C,2006csxs.book..157M,2019MNRAS.487..928S}. Thermal disc component dominating over the Comptonized component leads to High-soft state (HSS) or thermal state, during which the source commonly exhibits high disc temperature ($kT_{in}>1$ keV) and steep spectral index ($\Gamma>2$). Low Hard State (LHS) is seen when the high energy Comptonized component dominates over the disc component. Energy spectrum is characterised by disc component with low temperature ($kT_{in}<1$ keV) and dominant Comptonized component with $\Gamma<2$. Most of the persistent BH sources exhibit either HSS or LHS \citep{2002ApJ...578..357Z}, while outbursting sources depict two varieties of intermediate states between LHS and HSS i.e. Hard-Intermediate State (HIMS) and Soft- Intermediate State (SIMS) \citep{2001ApJS..132..377H,2005Ap&SS.300..107H,2005AIPC..797..197B,nandi2012accretion,2014AdSpR..54.1678R,2016MNRAS.462.1834R,2019MNRAS.487..928S, 2020MNRAS.497.1197B}.

BH sources also show different temporal properties such as presence of Quasi-periodic Oscillations (QPOs) in the PDS \citep{psaltis99,2001ApJ...561.1016M,2011BASI...39..409B,nandi2012accretion,2020arXiv200108758I} during different spectral states. The PDS in HSS is in the form of powerlaw noise component \citep{1997A&A...322..857B} with minimal variability and the QPOs are rarely observed in this state \citep{2001ApJS..132..377H,2005AIPC..797..197B,2006ARA&A..44...49R,2012A&A...542A..56N,2014AdSpR..54.1678R,2018JApA...39....5S}. Whereas during LHS, PDS contains low frequency QPO along with a flat top noise \citep{psaltis99, 2000MNRAS.318..361N, 2004astro.ph.10551V,nandi2012accretion,2016AN....337..398M,2020arXiv200108758I}. \par

LMC X-1 and LMC X-3 are persistent BH-XRBs located in the Large Magellanic cloud (LMC) at a distance of $\sim48.1$ kpc \citep{2009ApJ...697..573O}. Despite of its large distance, they can be observed in a broad X-ray band since minimal Galactic hydrogen column density $n_{H}$ is present along their line of sight. These sources are well known for the presence of persistent soft spectral state. While LMC X-3 has been found to undergo occasional transition into LHS, LMC X-1 has always been observed in HSS and never seem to have undergone any state transition. Most recently, broadband study of both these sources using \textit{AstroSat} observations \citep{2021MNRAS.501.5457B} confirmed the `extreme' soft nature whose energy spectra are dominated by thermal disc component. Soft-state spectrum of both sources are characterized using thermal disc blackbody component and the hard state spectra of LMC X-3 using high energy Comptonization component along with the disc having a spectral index $\Gamma\sim1.7$ \citep{2000ApJ...542L.127B,2001MNRAS.320..327W}. Temporal studies show that both LMC X-1 and LMC X-3 exhibits moderate variability in soft state with rms $\sim11.5\%$ and $\sim17\%$ respectively \citep{2021MNRAS.501.5457B} in wide energy band ($3.0-20$ keV). LMC X-3 in its hard state shows a high variability (rms$\sim40\%$) along with presence of QPO \citep{2000ApJ...542L.127B}. 
 \par
LMC X-1 is a rapidly spinning BH whose dimensionless spin parameter $a$ is estimated by continuum-fitting method \citep{1997ApJ...482L.155Z} to be $0.92^{+0.05}_{-0.07}$ \citep{2009ApJ...701.1076G} and  $0.97^{+0.02}_{-0.25}$ \citep{2012MNRAS.427.2552S} by Fe-line fitting method. LMC X-3 is estimated to have a low spin with $a=0.25^{+0.20}_{-0.29}$ \citep{2010ApJ...718L.117S}. In our previous work \cite{2021MNRAS.501.5457B}, using recent broadband \textit{AstroSat} observations, we have constrained the spin of both these sources using continuum-fitting method. Estimated spin value of LMC X-1 and LMC X-3 from this study are in the range $0.82-0.92$ and $0.22-0.41$ respectively. $M_{BH}$ is also constrained using same method whose value is found to be in the range of $7.64-10.00$ M$_{\odot}$ and $5.35-6.22$ M$_{\odot}$ for LMC X-1 and LMC X-3 respectively. These values are consistent with the dynamically estimated values which are $10.9\pm1.41$ M$_{\odot}$ \citep{2009ApJ...697..573O} for LMC X-1 and $6.98\pm0.56$ M$_{\odot}$ \citep{2001A&A...365L.273S,2014ApJ...794..154O} for LMC X-3. Inclination angle $i$ of the binary system constrained from the dynamical studies of the binary system is $36.38^{\circ}\pm1.92^{\circ}$ \citep{2009ApJ...697..573O} and $69.2^{\circ}\pm0.72^{\circ}$ \citep{2014ApJ...794..154O} respectively for LMC X-1 and LMC X-3. \par
In this work, we make use of \textit{NuSTAR} \citep{2013ApJ...770..103H}, \textit{NICER} \citep{2016SPIE.9905E..1HG}, and \textit{AstroSat} \citep{2001ASPC..251..512A} $-$ \textit{SXT} and \textit{LAXPC} observations to perform broadband spectral and timing analysis of the extragalactic BH-XRBs LMC X-1 and LMC X-3. \textit{MAXI} \citep{matsuoka2009maxi} data is used to plot the long term lightcurve and Hardness Ratio (HR) of both these sources. A comprehensive analysis is carried out by considering all the observations performed by these instruments during the period of $2014-2020$. We examine the variability of lightcurve, evolution of energy spectra and power spectra during different spectral states. We also study the broadband spectral and temporal properties of these sources by using the simultaneous observations of \textit{NICER} and \textit{NuSTAR}. Further, by modelling \textit{NICER} and \textit{NuSTAR} observations using relativistic accretion disc and reflection models, we constrain the spin and mass accretion rate of the BH sources. Results from the preliminary analysis of this work have been presented in the 43$^{rd}$ COSPAR Scientific Assembly \citep{cosparbhu}. \par 
Organization of this paper is as follows: In Section \ref{sec2}, we present the details of observations and data reduction. Spectral and timing analysis methods are discussed in Section \ref{sec3} and in Section \ref{sec4}, results from this analysis are presented. Finally in Section \ref{sec5}, we discuss the implication of the results obtained in the context of accretion disc dynamics and conclude the salient findings of this study.

\section{Observations and Data Reduction}
\label{sec2}
In this work, all the observations of LMC X-1 and LMC X-3 carried out during the \textit{NuSTAR}, \textit{AstroSat} and \textit{NICER} era are studied in detail. All the \textit{NICER}, \textit{NuSTAR} and \textit{AstroSat} observations which are considered in the present study are listed in Tables \ref{tab1}, \ref{tab2} and \ref{tab3}. We also consider the \textit{MAXI} lightcurve data observed during this period to study the long-term variability of the sources. 
\subsection{\textit{NuSTAR}}
\label{sec2.1}
\begin{figure*}
\centering
\includegraphics[width=14cm,height=7cm]{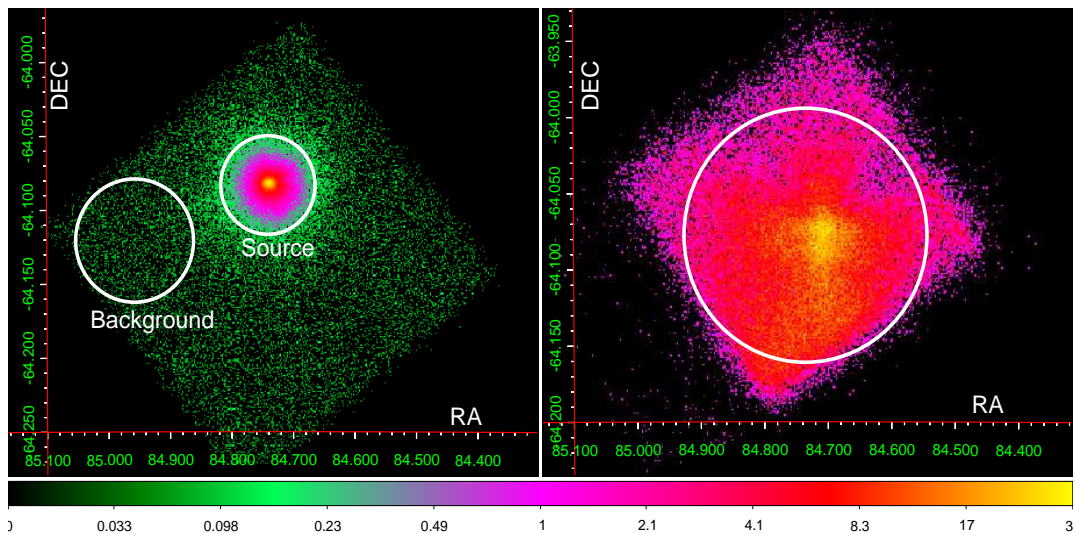}
\caption{\textit{NuSTAR-FPMA} (left) and \textit{AstroSat-SXT} (right) images of LMC X-3 obtained for the observation of MJD 58532 and MJD 58155 respectively. \textit{SXT} image is taken in FW mode. We consider a circular region of $2^{\prime}$ in the \textit{NuSTAR} image to extract source spectrum and lightcurve. Background files are extracted from a circular region away from the source. In \textit{SXT} image, the region of source selection is $4^{\prime}$. See text for details.}
\label{fig1}
\end{figure*}

 \textit{NuSTAR} consists of two Focal Plane Modules (FPM) referred to as \textit{FPMA} and \textit{FPMB} which operates in the energy band $3-79$ keV. For our analysis, observational data from both these instruments are used. \textit{NuSTAR} has observed LMC X-1 and LMC X-3 thrice during different epochs. Among these, the observation corresponding to observation ID \texttt{30402035005} of LMC X-3 is simultaneous with that of \textit{NICER} observation (\texttt{1101010136}). Following the standard guidelines (\footnotemark{\url{https://heasarc.gsfc.nasa.gov/docs/nustar/analysis/}}), we extracted the event file using \texttt{nupipeline} module. Source spectrum and lightcurve are obtained from a circular region of $2^{\prime}$ from \textit{FPMA} and \textit{FPMB} using \texttt{nuproducts}. The same module is used to obtain the redistribution matrix file (rmf) and ancillary response file (arf) for each observation. To generate the background file, larger source free region is used (Figure \ref{fig1}, left panel). Data grouping of energy spectra is carried out with 35 counts per bin.

\label{sec2.2}
\subsection{\textit{NICER}}
The \textit{Neutron star Interior Composition Explorer (NICER)} is an X-ray mission having X-ray Timing Instrument (\textit{XTI}) on-board, which operates in  $0.2 - 12$ keV energy band \citep{2016SPIE.9905E..1HG}. Excellent spectral and timing resolution of $6 < \frac{E}{\Delta E} < 80$ 
in 0.5 $-$ 8 keV and $< 300$ nsec respectively makes it an excellent instrument to study the spectral and timing properties of BH-XRBs. \par 
\textit{NICER} observed LMC X-1 during two occasions in 2018 and LMC X-3 has been observed at regular interval during the period of $2017 - 2020$. We processed these observations using \texttt{HEASARC v6.28} and \texttt{NICERDAS v7.0} and CALDB version \texttt{xti20200722} to obtain cleaned event files. Screening of data is done by using standard criteria of pointing offset $< 54^{\prime\prime}$, 
dark earth limb $> 30^{\circ}$, bright earth limb $> 40^{\circ}$ and outside of the South Atlantic Anomaly (SAA) region using \texttt{NIMAKETIME} and \texttt{NICERCLEAN} routines. Along with this condition, data from the noisy detectors 14 and 34 are removed using \texttt{fselect} command. Using \texttt{XSELECT}, we extract the energy spectra and lightcurve in $0.2-12$ keV energy band. Background is calculated using $3C50\_RGv5$ (\footnotemark{\url{ https://heasarc.gsfc.nasa.gov/docs/nicer/tools/n_icer_bkg_est_tools.html}}) model provided by the \textit{NICER} team  (see also \citealt{blessymaxi} under review). The background flux level is found to be less than $10\%$ during each observation in the entire $0.2-12$ keV energy range.  We use the latest standard response file provided by the \textit{NICER} team. The arf for 50 detectors are added to obtain the final arf. Spectral data are grouped with 25 counts per bin.
\subsection{\textit{AstroSat}}
\label{sec2.3}
\textit{Soft X-ray Telescope (SXT)} and \textit{Large Area X-ray Proportional Counter (LAXPC)} on-board \textit{AstroSat} simultaneously observes celestial sources in energy band of $0.3-8$ keV and $3-80$ keV respectively. The broadband coverage by these instruments along with the excellent spectral and timing capabilities provide good opportunity to study the BH sources more efficiently. \par
\textit{AstroSat} has observed LMC X-1 and LMC X-3 during 5 and 9 different epochs respectively. These data are obtained from the Indian Space Science Data Centre (ISSDC) archive (\footnotemark{\url{http://astrobrowse.issdc.gov.in/astro\_archive/archive}}). From a circular region of $4^{\prime}$ (Figure \ref{fig1}, right panel) and $10^{\prime}$ in the \textit{SXT} image of the sources in FW and PC mode respectively, the source spectra and lightcurves are extracted using \texttt{XSELECT V2.4g} \citep{2017JApA...38...29S}. The \texttt{sxtARFmodule} tool is used to generate the arf following the methodology suggested by the {\it SXT} team.
\textit{LAXPC} Level-1 data is converted to Level-2 using standard pipeline software (\footnotemark{\url{ https://www.tifr.res.in/~astrosat\_laxpc/LaxpcSoft.html}}). Standard background and response files provided by the instrument team are used (see also \citealt{2017caantialibration,2019MNRAS.487..928S,2020MNRAS.497.1197B,10.1093/mnras/staa3756,2021MNRAS.501.5457B} for details) for the analysis.

\subsection{\textit{MAXI}}
\label{sec2.4}
\begin{figure*}
\centering
\includegraphics[width=17cm,height=12cm]{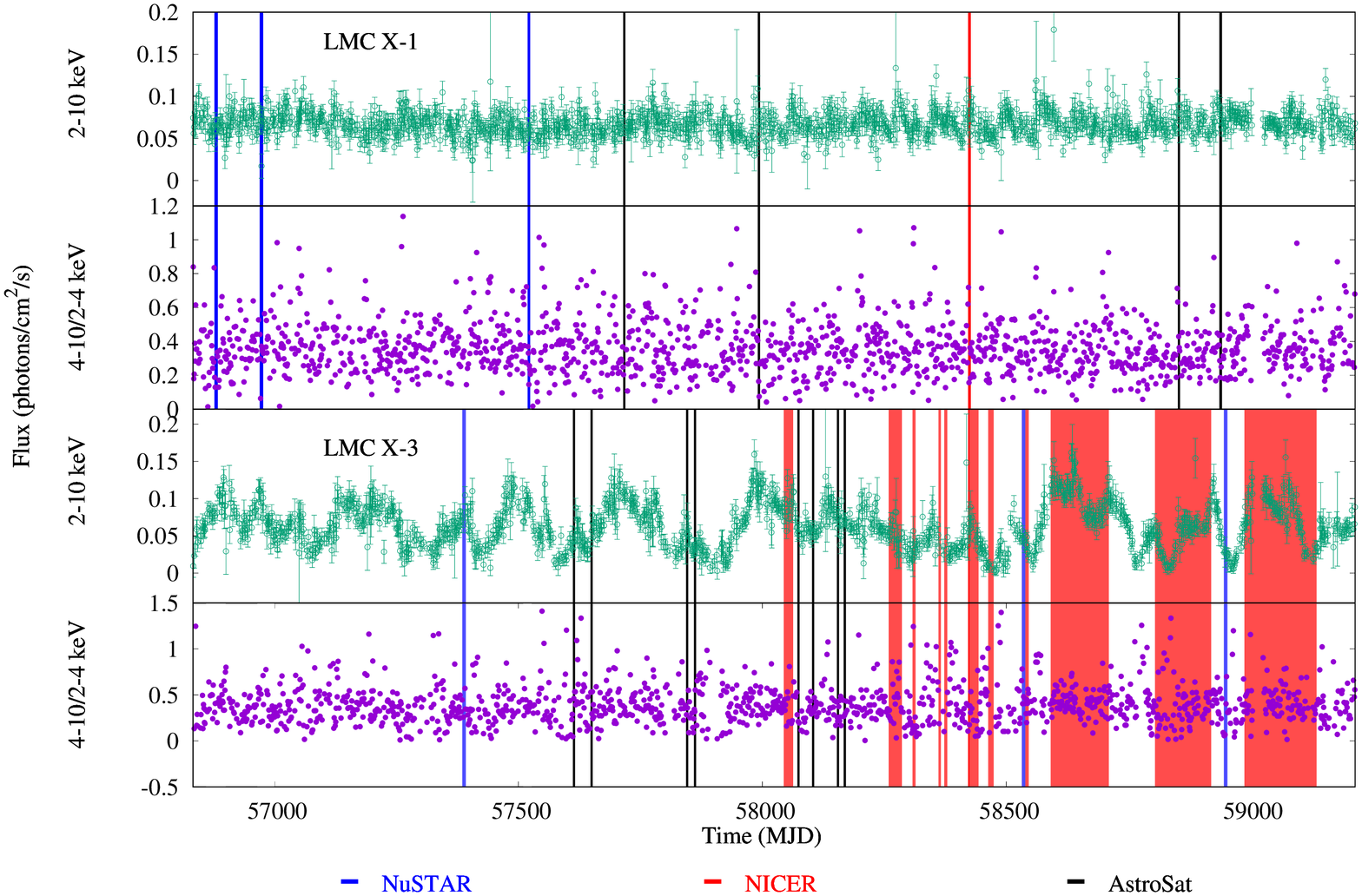}
\caption{Long-term (2014 $-$ 2020) \textit{MAXI} lightcurves and HR variations of LMC X-1  and LMC X-3 are shown in this figure. First panel represents the LMC X-1 lightcurve in $2-10$ keV, second panel shows the hardness ratio ($4-10$/$2-4$ keV) for LMC X-1. In the third panel, lightcurve of LMC X-3 in $2-10$ keV is shown along with the HR in panel 4. {\it NuSTAR} observations considered in this study are marked in blue lines, \textit{NICER} in red and \textit{AstroSat} in black. The red band indicates the \textit{NICER} observations performed in regular interval.}
\label{fig2}
\end{figure*}

The one day average lightcurve of LMC X-1 and LMC X-3 in the energy range of $2-10$ keV, $2-4$ keV and $4-10$ keV are obtained from the {\it MAXI} web-page \footnotemark{\url{http://maxi.riken.jp/top/lc.html}}. Lightcurve is plotted in $2-10$ keV for the period of MJD 56809 to MJD 59214 (2014$-$2020) to study the long-term variability (Figure \ref{fig2}). We also look into the HR by taking the ratio of flux in $4-10$ keV and $2-4$ keV ranges.

\section{Analysis and Modeling}
\label{sec3}
\subsection{Spectral Analysis}
\label{sec3.1}
\begin{figure*}
    \centering
    \includegraphics[height=8.5cm,angle=-90]{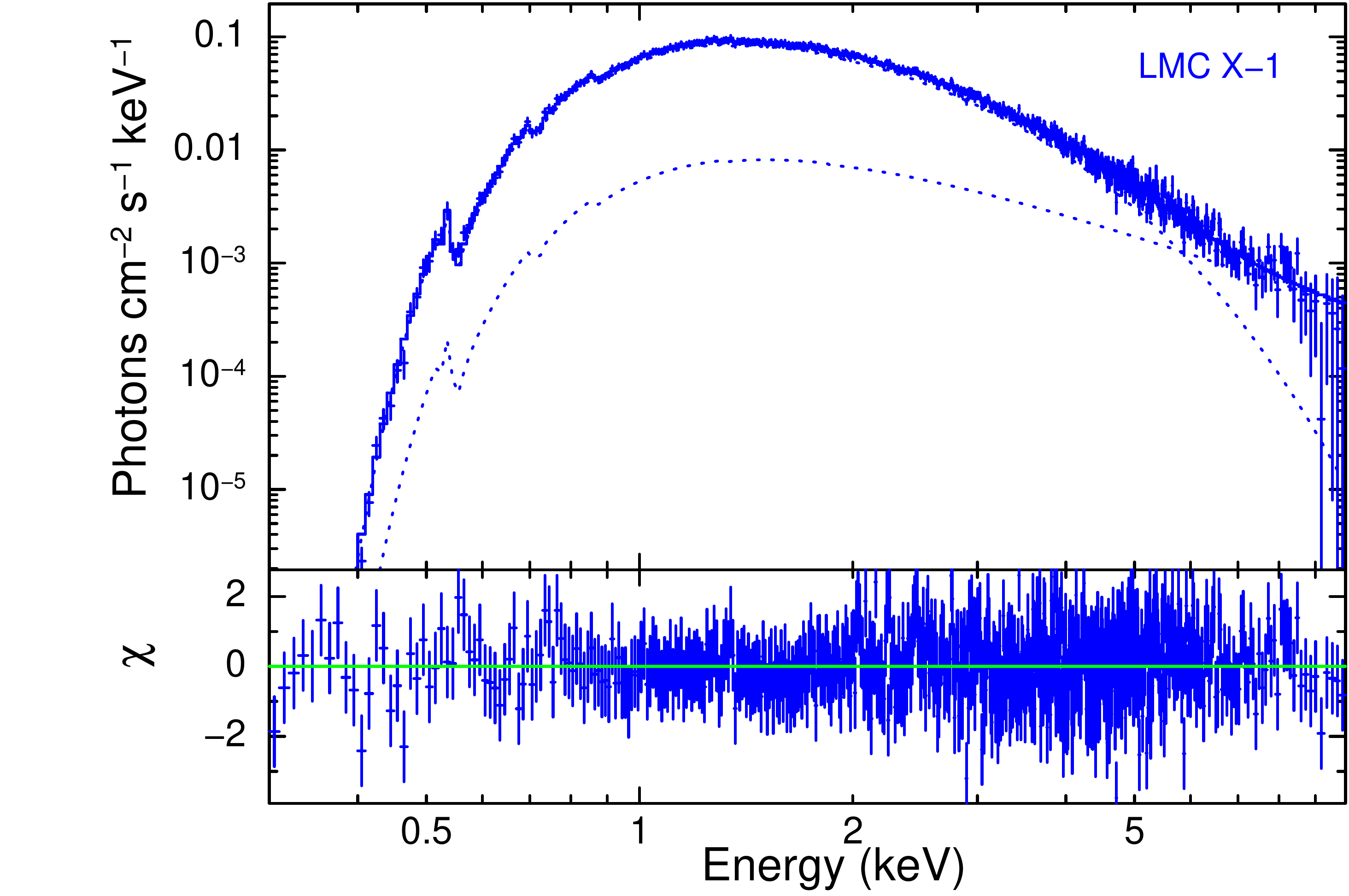}
    \includegraphics[height=8.5cm,angle=-90]{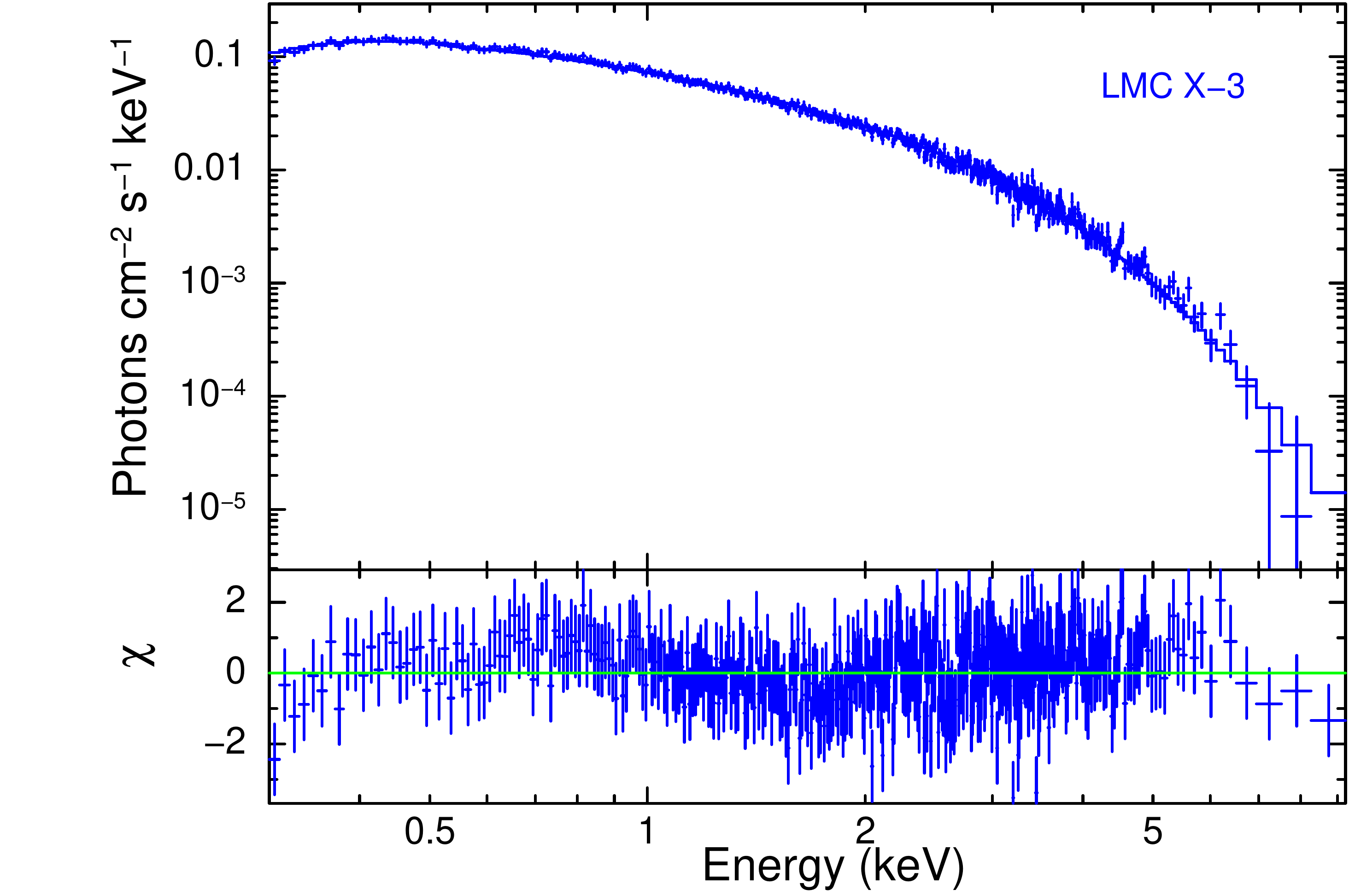}
    \caption{ The 0.3 $-$ 10 keV \textit{NICER} energy spectrum of LMC X-1 (left) and LMC X-3 (right) belonging to observations carried out on MJD 58422 and MJD 58373 respectively. Spectra are modelled using Model-1 by including systematic error of 5$\%<2$ keV and 1 $\%>$ 2 keV. Corresponding residuals from this fit are shown in bottom panel of both plots. }
    \label{fig3}
\end{figure*}

To perform the spectral analysis, \texttt{XSPEC v12.11.1} \citep{1996ASPC..101...17A} tool of \texttt{HEASoft v6.28} is used. Spectral data of LMC X-1 obtained by \textit{NICER} observations are modelled in the energy range of $0.3-10$ keV by grouping the spectra with 25 counts per bin. Large residues below 2 keV are found in most of the \textit{NICER} spectra which arises due to several instrumental uncertainties (\footnotemark{\url{https://heasarc.gsfc.nasa.gov/docs/nicer/analysis_threads/cal-recommend/}}). To account for this, a systematic error of 5\% is added to energy band of $0.3-2.0$ keV and $1\%$ above 2 keV following \cite{2020MNRAS.497.3896A}.  We begin the modelling of \textit{NICER} spectra of both LMC X-1 and LMC X-3 by using multi color disc blackbody model (\textit{diskbb}) and a power-law distribution model (\textit{powerlaw}). While this model combination gives a fairly good fit, in order to obtain the physical parameters of the sources, \textit{powerlaw} is replaced by \textit{nthcomp} \citep{1999MNRAS.309..561Z}. We also replace \textit{diskbb} with \textit{ezdiskbb} which assumes zero torque at the inner boundary and its normalization value is dependent on the spectral hardening factor \citep{2005ApJ...618..832Z}. Thus the model combination used to fit \textit{NICER} spectra of both sources is \textit{Tbabs(ezdiskbb+nthcomp)}  which we hereafter refer to as Model-1. In Figure \ref{fig3}, we demonstrate the Model-1 fitting to the \textit{NICER} spectrum of LMC X-1 (left panel) and LMC X-3 (right panel) and the respective residuals are shown in the bottom panel. Here, \textit{Tbabs} \citep{2000ApJ...542..914W} is used to account for the absorption of soft X-rays by interstellar medium. We use the elemental abundance values of the host galaxy while calculating the $n_H$ following \cite{hanke10} (see also \citealt{2021MNRAS.501.5457B}). \textit{ezdiskbb} was required for most of the considered observations to account for the emission from the accretion disc, while the \textit{nthcomp} is used to fit the high energy Comptonized component. During two observations of LMC X-3 i.e on MJD 58821 and MJD 58831, \textit{nthcomp} alone gives good fit without the requirement of \textit{ezdiskbb}. In all the fits where both disc and Comptonization component is present, seed photon temperature ($kT_{bb}$) of \textit{nthcomp} is tied to maximum disc temperature ($T_{max}$) of \textit{ezdiskbb}. We could not constrain the electron temperature $kT_{e}$ in any of the \textit{NICER} fits and thus we fix it to 20 keV as rest of the fit parameters are found to be insensitive to $kT_{e}>10$ keV. Not all energy spectra of LMC X-3 has Comptonized component and hence those spectra are fitted using \textit{ezdiskbb} alone. The unabsorbed source flux and disc flux contribution in $0.3-40$ keV are calculated from the \textit{NICER} energy spectra using \textit{cflux} model. Same model is used to derive the bolometric flux $F_{bol}$ in $0.1-50$ keV from which bolometric luminosity $L_{bol}$ is calculated by using the relation $L_{bol}=4 \pi D^{2}F_{bol}$ where $D$ is the distance to the source. 
$L_{bol}$ is used to estimate the $\dot{M}$ using the relation, $L_{bol}=\eta \dot{M} c^{2}$ where $\eta$ is the accretion efficiency whose value can be considered as $0.06$ for BHs with low spin and $0.3$ for highly spinning BHs and $c$ is the velocity of light. In order to express $\dot{M}$ in terms of $\dot{M}_{Edd}$, same relation is used to calculate $\dot{M}_{Edd}$ from the Eddington luminosity $L_{Edd}=1.3\times10^{38} \frac{M_{BH}}{M_{\odot}}$ erg s$^{-1}$. Average value of $M_{BH}$ obtained in \cite{2021MNRAS.501.5457B} is used to estimate L$_{Edd}$. 

\par 
The \textit{NuSTAR} energy spectra of both sources during all observations are found to have significant flux till $\sim40$ keV after which background flux dominates over the source flux resulting in poor statistics. Therefore all of the \textit{NuSTAR} spectral fits are carried out in the energy range $3.0-40$ keV. Spectra from both \textit{FPMA} and \textit{FPMB} are modeled by having an additional \textit{constant} parameter and allowing it to vary between the two data sets. The cross-normalization value of \textit{FPMB} is found to be within $5\%$ of the \textit{FPMA} in every fit which is in agreement with \cite{2015ApJS..220....8M}.  For all the \textit{NuSTAR} fits we could not constrain $n_{H}$ value and hence we fix it to the value obtained by \textit{NICER} spectral fits belonging to nearest observation. We tie $kT_{bb}$ to $T_{max}$ in all these fits and fix $kT_{e}$ to 20 keV as it could not be constrained. The presence of Fe K${\alpha}$ line at $\sim7$ keV is seen in the \textit{NuSTAR} spectra of LMC X-1. Therefore, to fit this emission line we initially use the \textit{Gauss} model. Although it gives an overall good fit, to incorporate the general relativistic effects, we replaced \textit{Gauss} by a more physical model \textit{laor} \citep{1991ApJ...376...90L} which describes the line profile obtained from the Kerr BHs. The line energy ($lineE$) and inner disc radius ($R_{in}$) are allowed to vary freely and the outer disc radius is set to default value of $400$ R$_{g}$ (where R$_{g}=GM/c^{2}$). Emissivity index $q$ is fixed at 3 which is the default value for standard disc and the system inclination angle is fixed to known values (see Section \ref{sec1}). However, we do not see Fe-emission feature in both \textit{NICER} and \textit{NuSTAR} spectra of LMC X-3.  
With the help of \textit{cflux} we calculate the $F_{bol}$ in $0.1-50$ keV and $f_{disc}$ in $0.3-40$ keV. $\frac{\dot{M}}{\dot{M}_{Edd}}$ is also calculated using the methodology mentioned before.\par
We model the source energy spectra obtained from \textit{AstroSat} observations by performing joint fitting of \textit{SXT} and \textit{LAXPC} in $0.5-20$ keV for LMC X-1 and $0.5-10$ keV for LMC X-3. Restriction of the spectral energy range is due to the unavailability of data with good statistics in high energy and background flux dominates over source flux. For Model-1 fitting, $kT_{bb}$ is tied to $T_{max}$ and $kT_{e}$ is fixed to 20 keV. From this fitting, we constrain the value of $n_{H}$, $T_{max}$ and $\Gamma$ of the sources during different observations. No Fe-line emission is seen in both sources during any of the \textit{AstroSat} observations. $F_{bol}$, $f_{disc}$ and accretion rate ratio are obtained using similar method as mentioned before.
\par
A simultaneous observation of LMC X-3 is carried out on MJD 58532 by \textit{NICER} and \textit{NuSTAR} and therefore broadband (0.3$-$40 keV) modelling of its spectrum is performed using Model-1. Fitting of three data sets i.e \textit{NICER}, \textit{FPMA} and \textit{FPMB} is carried out by using a \textit{constant} parameter which is fixed to 1.0 for \textit{FPMA} data set while it is allowed to vary freely for \textit{NICER} and \textit{FPMB}. The cross-normalization value of \textit{NICER} is found to be close to 1. Tying $kT_{bb}$ to $T_{max}$ in this joint fit results in large residue below 0.5 keV which is not seen when these two temperatures are left untied. Therefore the fitting is carried out by allowing $T_{max}$ and $kT_{bb}$ to vary independently. 
%We obtained an acceptable fit ($\chi^{2}/dof= 1231.8/1268$) by letting the \textit{ezdiskbb} parameters of {\it NICER} to be varying and tie the respective in \textit{NuSTAR} to that of \textit{NICER}. In a similar way, the {\it nthcomp} parameters of NuSTAR are kept free, to which the corresponding parameters from {\it NICER} are tied to.

\subsection{Continuum-fitting and Fe-line fitting Method}
\label{sec3.2}
The physical parameters of BH sources such as spin and mass accretion rate can be estimated by fitting the thermal X-ray continuum in the energy spectrum of the BH \citep{1997ApJ...482L.155Z,2014SSRv..183..295M,2009ApJ...701.1076G,2010ApJ...718L.117S,2021MNRAS.501.5457B} and this is known as continuum-fitting method. Since soft state energy spectra has high accretion luminosity, they are most suitable to carry out the continuum fitting method. We have presented the estimation of physical parameters of both LMC X-1 and LMC X-3 by implementing this method using \textit{AstroSat} observations in \cite{2021MNRAS.501.5457B}. In this work, we adapt the same method to estimate the physical parameters by using  \textit{NICER} and \textit{NuSTAR} observations. For this purpose, we model the \textit{NICER} and \textit{NuSTAR} energy spectra in $0.3-10$ keV and $3-40$ keV respectively using a relativistic accretion disc model \textit{kerrbb} \citep{2005ApJS..157..335L}.
The model combination \textit{Tbabs(simpl $\otimes$ kerrbb)} is used to constrain the mass accretion rate ($\dot{M}$) and spin ($a$) of the BHs which we refer to as Model-2 throughout the paper. Here, \textit{simpl} \citep{2009PASP..121.1279S} models the Comptonization of soft photons which is required only in \textit{NuSTAR} spectra because of absence of Comptonization component in all the soft state \textit{NICER} spectra. An additional \textit{laor} model is used in all the \textit{NuSTAR} fits of LMC X-1 due to the presence of Fe line emission. \textit{kerrbb} fits the low energy disc continuum by including the relativistic effects and thereby estimates physical parameters of the source.For these fits, distance to binary system and the inclination angle fixed to the dynamically estimated values (see Section \ref{sec1}) and normalization is fixed to 1. Mass is fixed to the averaged value obtained by \cite{2021MNRAS.501.5457B} which is $8.82$ M$_{\odot}$ and $6.48$ M$_{\odot}$ for LMC X-1 and LMC X-3 respectively. The spectral hardening factor is fixed at 1.55 for LMC X-1 and 1.7 for LMC X-3 in all the fits (see also \citealt{2021MNRAS.501.5457B} for more details). Spectral parameters obtained from these fits are listed in Table \ref{tab4}. Errors of the parameters $a$ and $\dot{M}$ are calculated using the Markov Chain Monte Carlo (MCMC) method with Goodman-Weare algorithm in \texttt{XSPEC} by setting walkers parameter to 8 and length of chain to 10000 along with a burn length of 1000 (see also \citealt{2020MNRAS.499.5891S}, \citealt{2021MNRAS.501.5457B} and \citealt{2021MNRAS.tmp.2029K}).
\par
\begin{figure}
    \centering
    \includegraphics[height=9cm,angle=-90]{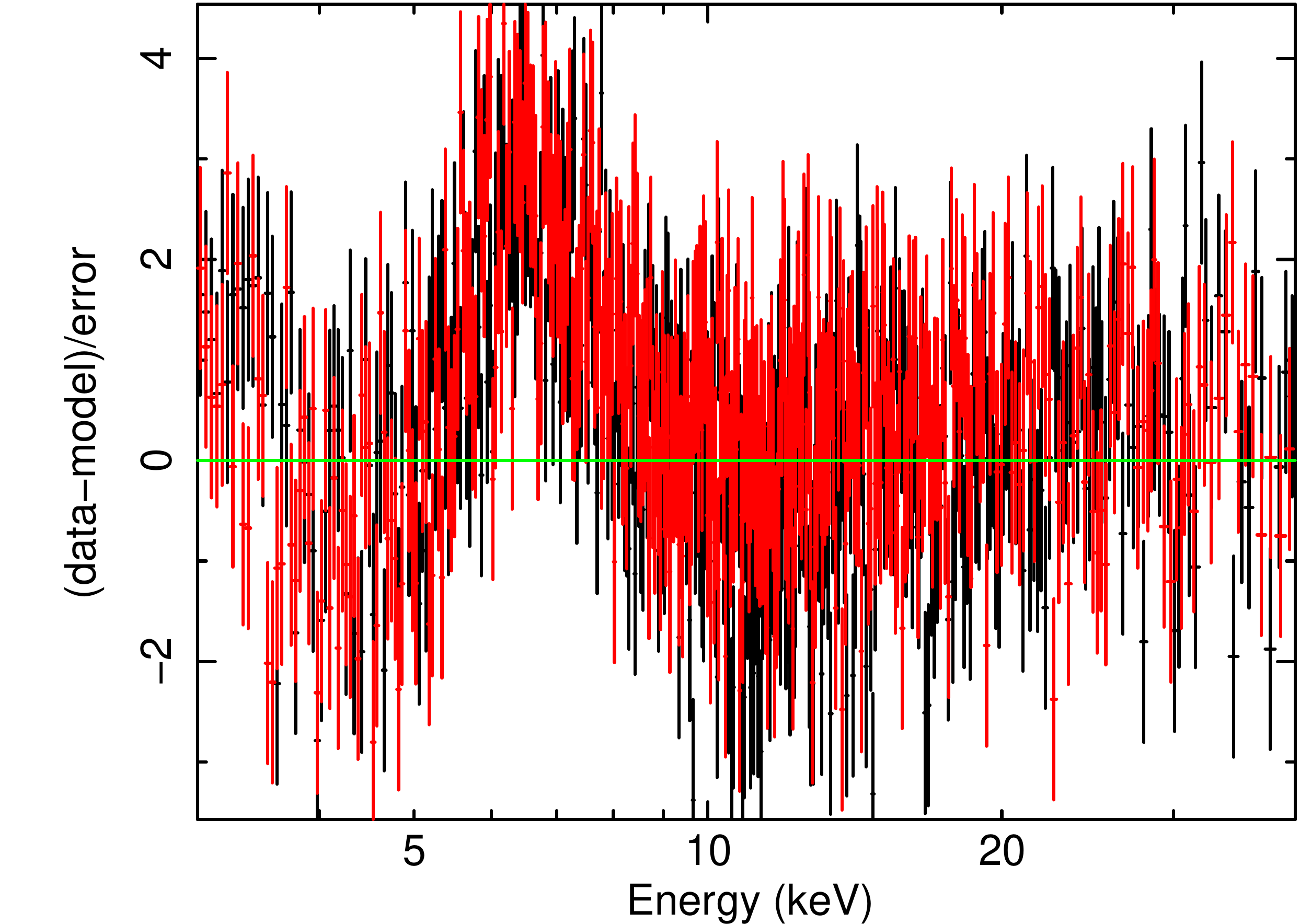}
    \caption{Residual of the Model-3 fit to the \textit{NuSTAR} spectra of LMC X-1 belonging to MJD 58946 without \textit{relline} model. The presence of residue in the 6$-$7 keV energy range shows the Fe-line emission profile.} 
    \label{fig4}
\end{figure}

In order to check for the possible Compton reflection feature present in the 15$-$30 keV spectrum, we attempt to model the \textit{AstroSat} and \textit{NuSTAR} spectra of both sources using the relativistic reflection model \textit{relxill} \citep{2014ApJ...782...76G} along with Model-2. The fit gives a reasonable $\chi^{2}_{red}$ however we could not constrain the individual parameters of \textit{relxill} in both sources. In Figure \ref{fig4}, we plot the residual of the continuum fitted \textit{NuSTAR} spectrum of LMC X-1 (MJD 58946) which depicts the presence of Fe-line feature in its spectrum. We model this using a reflection line model \textit{relline} \citep{2010MNRAS.409.1534D} to all three \textit{NuSTAR} spectra of LMC X-1. The model combination \textit{Tbabs(simpl $\otimes$ kerrbb+relline)} shall be referred as Model-3. Here, emissivity indices q$_{1}$ and q$_{2}$ are kept fixed at their default value of 3 and the break radius to 15 R$_{g}$. We fix the $i$ to $36.38^{\circ}$ and tie the spin parameter $a_{*}$ of \textit{relline} to parameter $a$ of \textit{kerrbb} by keeping $R_{in}$ free while outer disc radius $R_{out}$ is fixed at the default value of 400 R$_{g}$. In the next step we fix the $R_{in}$ to this obtained value and untied the parameters $a$ and $a_{*}$. This allowed us to constrain the spin by using continuum-fitting method (using \textit{kerrbb}) as well as from  Fe-line fitting method (using \textit{relline}).
\par

\subsection{Temporal Analysis}
\label{sec3.3}
Long-term lightcurve data of LMC X-1 and LMC X-3 in the energy band $2-4$ keV and $4-10$ keV with 1 day time-bin are obtained from \textit{MAXI} observations. HR is calculated by taking the ratio of flux in two energy bands i.e $(4-10)/(2-4)$ keV. Variability of the lightcurve in 2$-$10 keV is estimated by calculating the fractional variance $F_{var}$ for the period 2014$-$2020 following the method mentioned by \cite{2003MNRAS.345.1271V} (see \citealt{2021MNRAS.501.5457B} for details). \par 
The timing properties of both sources studied using \textit{AstroSat} observations have been already presented in \cite{2021MNRAS.501.5457B}. In this work, we make use of \textit{NICER} and \textit{NuSTAR} lightcurves for temporal studies. For this purpose, we obtained \textit{NICER} and \textit{NuSTAR} lightcurves in the energy range of $0.3-10$ keV and $3-79$ keV respectively with a time bin of 50 ms. We generated PDS for each of these observations in the frequency range $0.002-10$ Hz by taking the average of PDS generated for intervals of 8192 bins. Geometric re-binning of PDS by a factor of 1.05 is carried out \citep{1990A&A...230..103B,1999ApJ...510..874N} in frequency space. Normalized PDS are corrected for Poisson noise.\par 
We model the PDS using \textit{powerlaw} model in rms space \citep{2002ApJ...572..392B}. The rms values of PDS are computed for the $0.002-10$ Hz frequency range using rectangle rule integration method (see also \cite{riemann_1867,2018Ap&SS.363..189D} and references therein). 

\section{Results}
\label{sec4}
\subsection{Spectral Properties}
\subsubsection{LMC X-1}
\label{sec4.1.1}
\begin{figure}
\includegraphics[height=9cm,angle=-90]{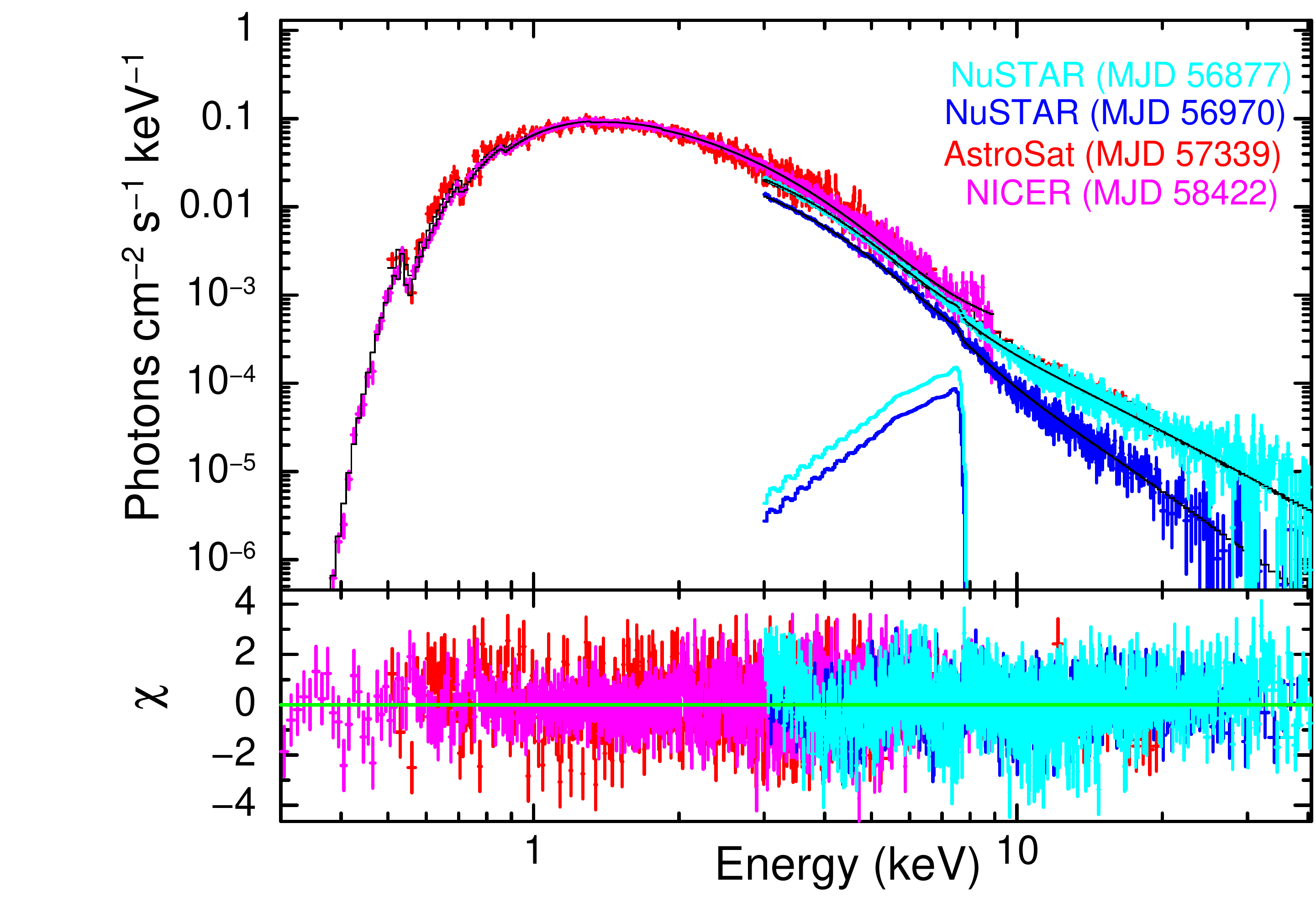}
\caption{Unfolded spectra of LMC X-1 plotted using \textit{NuSTAR} observations performed on MJD 56877 and MJD 56970, \textit{AstroSat} on MJD 57339 and \textit{NICER} on MJD 58422. Spectral fitting is carried out using Model-1. For the plot, spectra are re-binned to have 50 counts in each bin.} See text for more details.
\label{fig5}
\end{figure}

As described in Section \ref{sec3.1}, the \textit{NICER}, \textit{NuSTAR} and \textit{AstroSat} energy spectra of LMC X-1 are modelled using Model-1 in the energy ranges of $0.3-10$ keV, $3-40$ keV and $0.5-20$ keV respectively. Fitting the two \textit{NICER} energy spectra belonging to two observations using Model-1, resulted in $n_{H}$ value of $1.25-1.27\times10^{22}$ atoms cm$^{-2}$. The energy spectra of both observations have a thermal disc blackbody component with similar disc temperature $T_{max}$ of $0.76\pm0.02$ and $0.78\pm0.01$ keV and normalization $N_{ezd}$ of $22.97^{+1.11}_{-0.13}$ and $19.10^{+0.77}_{-0.95}$. The source spectrum has spectral indices $\Gamma=2.06-2.12$ and unabsorbed source flux of $\sim1.02\times10^{-9}$ erg cm$^{-2}$ sec$^{-1}$. The disc flux is found to be dominant with $f_{disc}>80\%$ along with high $\dot{M}$ of $\sim0.24$ $\dot{M}_{Edd}$.  The spectral parameters of LMC X-1 obtained from the \textit{NICER} fits are listed in Table \ref{tab1}. \\
Similar properties are also seen by modelling the \textit{NuSTAR} spectra. Fitting Model-1 to all of the \textit{NuSTAR} spectra resulted in $T_{max}$ in the range of $0.80-0.87$ keV and $N_{ezd}$ in $6.98-14.34$, along with high $\Gamma$ of $2.68-3.53$. In addition, \textit{NuSTAR} energy spectra also consist of Fe emission line within $7.11-7.18$ keV which is accounted by the \textit{laor} model, which resulted in $R_{in}\sim2.58$ R$_{g}$ in all three \textit{NuSTAR} observations. Unabsorbed source flux is within $2.89 - 4.69\times10^{-10}$ erg cm$^{-2}$ sec$^{-1}$ and $\dot{M}$ within $0.07-0.11$ $\dot{M}_{Edd}$. \textit{NuSTAR} spectra are also found to have dominant disc flux contribution with $f_{disc}>80\%$. The resultant spectral parameters of LMC X-1 from these \textit{NuSTAR} fits are given in Table \ref{tab2}. \\
Consistency of the spectral parameters are further confirmed by spectral results from \textit{AstroSat} observations. We present the obtained spectral parameters from Model-1 fitting to \textit{AstroSat} energy spectra in Table \ref{tab3}. $T_{max}$, $N_{ezd}$ and $\Gamma$ are found to be within $0.83-0.90$ keV, $5.89-11.05$ and $2.53-4.08$ respectively. During these observations,  $f_{disc}$ is found to be $>89\%$ and $F_{bol}$ within $2.95-3.92 \times10^{-10}$ erg cm$^{-2}$ s$^{-1}$ which corresponds to $\dot{M}$ of 0.07$-$0.09 $\dot{M}_{Edd}$. \\
In Figure \ref{fig5}, we have shown the unfolded energy spectra of LMC X-1 plotted using \textit{AstroSat}, \textit{NuSTAR} and \textit{NICER} observations that are modelled with Model-1. Presence of Fe emission line in the \textit{NuSTAR} spectra is also shown.  All of these spectra observed during different epochs belongs to soft spectral state.  \par 

\subsubsection{LMC X-3}
\label{sec4.1.2}
We model the \textit{NICER}, \textit{NuSTAR} and \textit{AstroSat} energy spectra of LMC X-3 in the energy range $0.3-10$ keV, $3.0-40.0$ keV and $0.5-10$ keV respectively using Model-1 as explained in Section \ref{sec3.1}. \textit{NICER} spectral fits resulted in $n_H$ value of $0.05-0.08 \times 10^{22}$ cm$^{-2}$ and, from \textit{AstroSat} fits it is found to be within $0.02-0.05 \times 10^{22}$ cm$^{-2}$. From the spectral analysis it is clear that unlike LMC X-1, LMC X-3 exhibits a spectral variability. While source exhibits consistent spectral behaviour during all the \textit{AstroSat} observations, \textit{NICER} and \textit{NuSTAR} energy spectra are found to have evolved throughout the observations. In Figure \ref{fig7} we have shown the unfolded \textit{NuSTAR} energy spectra of LMC X-3 in the top panel that is modelled using Model-1 and its residual is plotted in the bottom panel. The three energy spectra are observed during different epochs i.e on MJD 57385, 58532 and 58946. Variation of soft and hard X-ray flux as well as the spectral pattern can be seen in the plot which provides clear indication of source spectral evolution. Thus by modelling these as well the \textit{NICER} and \textit{AstroSat} energy spectra, we found that LMC X-3 occupies three distinct spectral states by undergoing state transition on several occasions within the period of 2015$-$2020. We list the spectral parameter values obtained from these fits in Tables \ref{tab1}, \ref{tab2} and \ref{tab3}. Source properties during different spectral states are as follows.
\begin{itemize}
    \item{Soft State:
    The first observation of LMC X-3 carried out by \textit{NuSTAR} is on MJD 57385 and it shows $T_{max}=1.10\pm0.01$ keV and $N_{ezd}=5.25\pm0.09$ along with $\Gamma=2.27\pm0.11$. $F_{bol}$ of $6.2\times10^{-10}$ erg cm$^{-2}$ sec$^{-1}$ is obtained for this observation along with $f_{disc}\sim95\%$. Derived bolometric luminosity translates into $\dot{M}$ of $\sim0.20$ $\dot{M_{Edd}}$ during this \textit{NuSTAR} observation. \par
    During the \textit{NICER} observations carried out over the period of MJD 58044 $-$ MJD 58443, $T_{max}$ remained constant with a value of $\sim1$ keV and $N_{ezd}$ is found to be in the range of $5.22-7.87$. High energy Comptonization component is absent in all these \textit{NICER} spectra making thermal accretion disc as a sole contributor to the total flux. Large flux and $\dot{M}$ are obtained for these \textit{NICER} observations having values of $5\times10^{-10} - 1.3\times10^{-9}$ erg cm$^{-2}$ sec$^{-1}$ and $0.16-0.42$ $\dot{M}_{Edd}$ respectively. Similar spectral behaviour is also found in the \textit{NICER} observations carried out for a brief period during MJD 58591 $-$ MJD 58805, and again from MJD 58878 until the latest observation done on MJD 59136.\par 
    Similar behavior is found from \textit{AstroSat} spectral analysis as well where \textit{nthcomp} component is absent in all observations. During different observations, $T_{max}$ is found to be $\sim1$ keV and $N_{ezd}$ within $3.35-4.55$. Source flux is contributed from thermal component alone whose value is found to be within $1.78-5.34 \times10^{-10}$ erg cm$^{-2}$ s$^{-1}$. $\dot{M}$ during these observations is estimated to be in the range $0.06-0.17$ $\dot{M}_{Edd}$. \par 
    These spectral characteristics found by \textit{NICER}, \textit{NuSTAR} and \textit{AstroSat} observations are similar to the soft state properties of LMC X-3 found by \textit{RXTE} \citep{2001MNRAS.320..316N,2001MNRAS.320..327W,2007ApJ...669.1138S}. }
    \item{Intermediate State: 
    Spectral analysis of \textit{NuSTAR} observations carried out on MJD 58532 and MJD 58946 resulted in slightly different parameter values when compared to the soft state spectra i.e  $T_{max}=0.88-1.00$ keV, $N_{ezd}=3.62-4.44$ 
    and $\Gamma=2.12-2.35$ are obtained from these energy spectra. A reduction in
    flux to $1.6-4.35\times10^{-10}$ erg cm$^{-2}$ sec$^{-1}$, $\dot{M}$ to $\sim0.09$ $\dot{M}_{Edd}$ and $f_{disc}$ to $69-72\%$ are observed, which are clear deviations from the soft state spectral properties. \par
    \textit{NICER} energy spectra observed on MJD 58463, MJD 58532 and for a brief period from MJD 58821 to MJD 58831 also show a decrease in $T_{max}$ and $\Gamma$ to $0.38-0.87$ keV, $2.02-2.36$ respectively and increase in $N_{ezd}$ to $6.39-23.83$. Similar to the characteristics explained for the \textit{NuSTAR} spectrum, the value of $F_{bol}$, $\dot{M}$ and $f_{disc}$ are found to have decreased during the {\it NICER} observations also to $1.2-4.2\times10^{-10}$ erg cm$^{-2}$ sec$^{-1}$, $\sim0.09$ $\dot{M}_{Edd}$ and $47-73\%$ respectively.
    \par
    During two \textit{NICER} observations i.e on MJD 58821 and MJD 58831, where spectrum is described by only \textit{nthcomp}, $F_{bol}$ is found to be $2.85\times10^{-10}$ erg cm$^{-2}$ sec$^{-1}$ and $1.14\times10^{-10}$ erg cm$^{-2}$ sec$^{-1}$ which is close to the flux during the above mentioned \textit{NICER} and \textit{NuSTAR} observations. \textit{nthcomp} fit results in $kT_{bb}$ of $0.51\pm0.01$ keV and $0.32\pm0.01$ keV and $\Gamma$ of $3.53\pm0.13$ and $2.8\pm0.08$ during the two observations. The broadband fit of simultaneously observed \textit{NICER} and \textit{NuSTAR} spectra on MJD 58532 resulted in $T_{max}=1.00\pm0.01$ keV, $N_{ezd}=4.39\pm0.10$, $kT_{bb}=0.43\pm0.03$ and $\Gamma=2.43\pm0.04$.  
     \par
    Spectral results from the analysis of these \textit{NICER} and \textit{NuSTAR} observations are consistent with the spectral properties of transient Galactic BHs in their intermediate state \citep{1997ApJ...479..926M,2006ARA&A..44...49R,2012A&A...542A..56N}}
    \item{Hard State: \textit{NICER} observation performed on MJD 58470 is found to have a cooler disc with $T_{max}=0.15\pm0.01$ keV. $N_{ezd}$ is found to have increased to $135.62^{+62.44}_{-42.62}$ and $\Gamma$ decreased to $1.6\pm0.07$. A low flux of $\sim2.41\times10^{-11}$ erg cm$^{-2}$ sec$^{-1}$ with $f_{disc}$ of only $26\%$ is obtained. Accretion rate during this observation is found to have decreased to $<0.01$ $\dot{M}_{Edd}$. Such behaviour is found only in a single observation by \textit{NICER}. Similar properties of LMC X-3 has been previously found by \textit{RXTE} observations of LMC X-3 that shows a significant drop in flux and has dominant Comptonization component in its energy spectrum and it has been classified as belonging to hard state  \citep{2000ApJ...542L.127B,2001MNRAS.320..327W}.  }
\end{itemize}
 In Figure \ref{fig7}, the unfolded energy spectrum of LMC X-3 during different spectral states are plotted using the observations from different instruments. Observations carried out on MJD 57650 represents soft state as observed by \textit{AstroSat} (red). MJD 58263 belongs to \textit{NICER} (magenta) soft state observation, hard state spectrum is observed on MJD 58470 by \textit{NICER} (orange) and source intermediate state is represented by MJD 58532 observation (blue) by \textit{NICER} and \textit{NuSTAR}.
 Clear distinction between different spectral states is evident with hard state spectrum having low flux and soft state having relatively high flux.\par
\begin{figure}
\includegraphics[height=9cm,angle=-90]{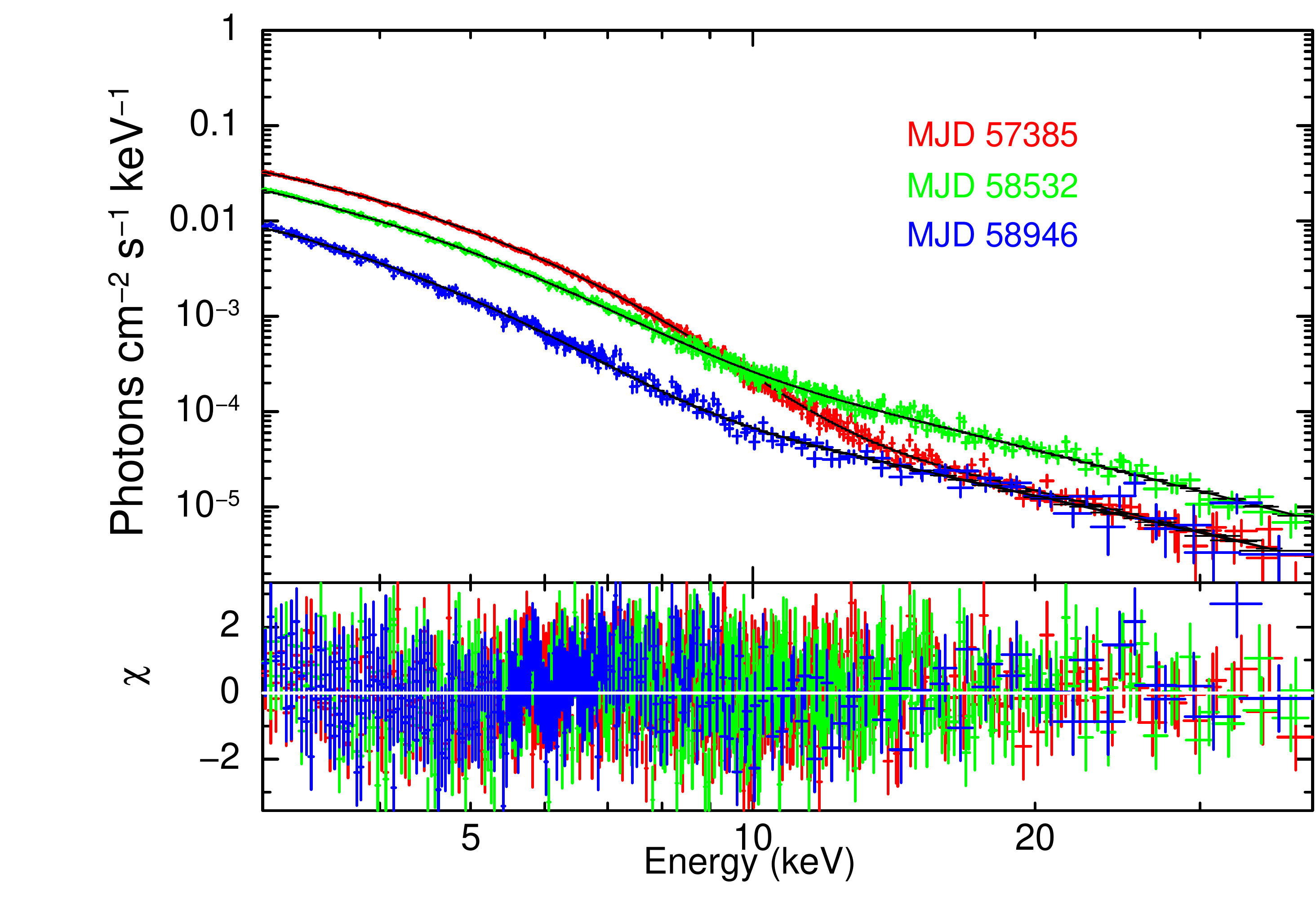}
\caption{\textit{NuSTAR} unfolded energy spectra of LMC X-3 belonging to observations carried out on MJD 57385, MJD 58532 and MJD 58946. Fitting is carried out using Model-1 in 3.0$-$40 keV. Spectra are re-binned to have 50 counts/bin for plotting purpose.}
 \label{fig6}
 \end{figure}
\begin{figure}
\includegraphics[height=9cm,angle=-90]{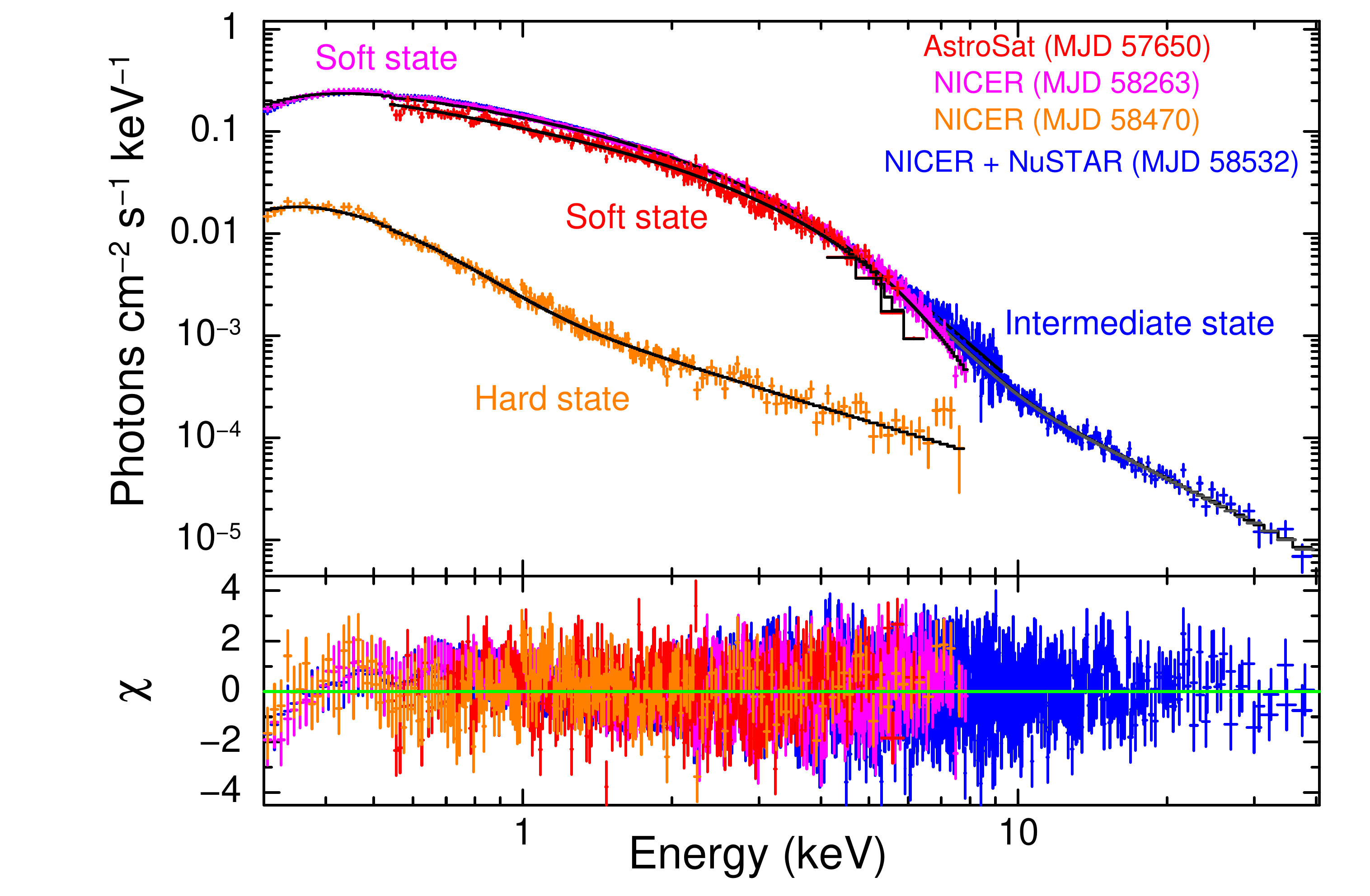}
\caption{Unfolded energy spectra of LMC X-3 plotted for observations done on  MJD 57650, 58263, 58470 and 58532 using \textit{AstroSat}, \textit{NICER} and \textit{NuSTAR} which represents the different spectral states. Modelling is carried out using Model-1. Each spectrum is re-binned to have 50 counts per bin for plotting purpose.} See text for more details.
 \label{fig7}
 \end{figure}
\begin{figure}
\includegraphics[width=9cm]{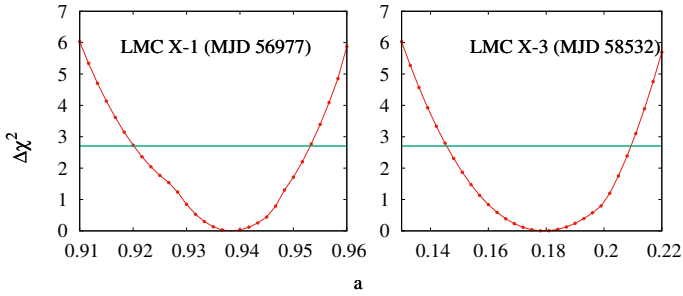}
\caption{Statistical variation of $\Delta\chi^{2}$ plotted at stepping values of BH spin for LMC X-1 (left) and LMC X-3 (right). The values are obtained from Model-2 fitting to the \textit{NuSTAR} energy spectra belonging to observations carried out on MJD 56977 and MJD 58532 for LMC X-1 and LMC X-3 respectively. The horizontal line denotes the $90\%$ confidence limit.}
\label{fig8}
\end{figure}
 
\begin{table*}
\caption{Tabulated \textit{NICER} observations of LMC X-1 and LMC X-3. The spectral parameters are obtained from the Model-1 (\textit{Tbabs(ezdiskbb+nthcomp)}) fitting to \textit{NICER} energy spectra in $0.3-10$ keV energy range. Errors for all the parameters are calculated with $90\%$ confidence. }
\label{tab1}
\begin{tabular}{lllllllll} \hline
ObsID      & \multicolumn{1}{c}{Date} & \multicolumn{1}{c}{MJD} & \multicolumn{1}{c}{nH}  & \multicolumn{1}{c}{$T_{max}$/$kT_{bb}$}     & $N_{ezd}$                   & $\Gamma$                     & \multicolumn{1}{c}{$F_{bol}$}  & $\chi^{2}/dof$                  \\
 & 
& & \multicolumn{1}{c}{\footnotesize{$\times10^{22}$}} &  & & & \multicolumn{1}{c}{\footnotesize{ $\times10^{-10}$ }} & \\
& 
& & \multicolumn{1}{c}{\footnotesize{( cm$^{-2}$})}& \multicolumn{1}{c}{\footnotesize{(keV)}} & & &  \footnotesize{( erg cm}$^{-2}$ s$^{-1}$) & \\
\hline
\multicolumn{9}{c}{LMC X-1}                                                                                                                   \\ \hline
1100070101 & 2020-10-31 & 58422 & $1.27\pm0.02$ & $0.76\pm0.02$  & $22.97^{+1.11}_{-1.28}$ & $2.06^{+0.79}_{-0.72}$    & $10.2$ & $518.78/589$         \\
1100070102 & 2020-11-01 & 58423 & $1.25\pm0.01$          & $0.78\pm0.01$  & $19.10^{+0.77}_{-0.95}$  & $2.12^{+0.38}_{-0.35}$   & $10.2$ & $597.21/653$     \\ \hline  
\multicolumn{9}{c}{LMC X-3}                                                                                                                   \\ \hline
1101010102 & 2017-10-18               & 58044                   & $0.03\pm0.01$ & $1.11\pm0.01$          & $6.01^{+0.26}_{-0.25}$    & $-$  & $13.2$  &435.16/483                      \\
1101010103 & 2017-10-19               & 58045                   & $0.04\pm0.01$ & $1.09\pm0.01$          & $6.26^{+0.21}_{-0.02}$           & $-$ & $12.8$ &  582.86/769      \\
1101010107 & 2017-10-27               & 58053                   & $0.04\pm0.01$  & $1.12\pm0.01$          & $5.69\pm0.06$    &$-$  & $11.2$ & 719.37/859                    \\
1101010117 & 2018-05-21               & 58259                   & $0.04\pm0.01$ & $0.99\pm0.01$          & $6.32\pm{0.16}$    &$-$ & $7.71$ & 540.97/513                     \\
1101010118 & 2018-05-25               & 58263                   & $0.04\pm0.01$ & $1.00\pm0.01$          & $6.01\pm0.11$    &$-$  & $7.50$ & 508.01/654                       \\
1101010119 & 2018-05-27               & 58265                   & $0.03\pm0.01$ & $0.98\pm0.01$          & $6.56\pm0.15$   &$-$  & $7.85$ & 573.89/606                       \\
1101010120 & 2018-05-28               & 58266                   & $0.04\pm0.01$ & $0.98\pm0.01$          & $6.56^{+0.18}_{-0.17}$     & $-$ & $7.66$ & 495.87/548                       \\
1101010121 & 2018-05-31               & 58269                   & $0.04\pm0.01$ & $0.97\pm0.01$          & $6.59\pm0.21$   &$-$ & $7.28$ & 420.64/506                       \\
1101010122 & 2018-06-01               & 58270                   & $0.04\pm0.01$ & $0.96\pm0.01$ & $6.43\pm0.21$    & $-$  &$6.90$& 517.42/478                       \\
1101010123 & 2018-06-02               & 58271                   & $0.04\pm0.01$ & $0.97\pm0.01$          & $6.33\pm0.12$   &$-$  &$6.91$& 566.64/598                       \\
1101010125 & 2018-06-17               & 58286                   & $0.04\pm0.01$ & $0.83\pm0.01$          & $7.87^{+0.42}_{-0.40}$   & $-$ & $4.76$ & 330.41/345                        \\
1101010126 & 2018-07-09               & 58308                   & $0.03\pm0.01$ & $0.86\pm0.01$          & $6.31\pm0.17$    & $-$  & $4.31$ & 511.99/522                       \\
1101010128 & 2018-08-31               & 58361                   & $0.04\pm0.01$ & $0.87\pm0.02$ & $6.97^{+0.49}_{-0.46}$   &$-$  & $4.95$ & 294.66/306                       \\
1101010129 & 2018-09-12               & 58373                   & $0.03\pm0.01$ & $0.76\pm0.01$          & $7.68\pm0.30$    &$-$ & $3.17$ & 347.33/401                       \\
1101010130 & 2018-11-03               & 58425                   & $0.04\pm0.01$ & $0.96\pm0.01$          & $6.73\pm0.12$   & $-$   & $7.17$ & 649.27/599                       \\
1101010131 & 2018-11-17               & 58439                   & $0.04\pm0.01$ & $0.94\pm0.01$          & $6.33\pm0.20$    &$-$ & $6.33$ &466.06/508                       \\
1101010132 & 2018-11-21               & 58443                   & $0.04\pm0.01$ & $0.94\pm0.01$          & $6.31\pm0.17$   &$-$ & $6.13$ & 530.07/554                      \\
1101010133 & 2018-12-11               & 58463                   & $0.05\pm0.01$ & $0.38\pm0.01$          & $23.22^{+2.77}_{-2.48}$   & $2.36^{+0.05}_{-0.04}$   &  $1.50$ & 253.92/341                                \\
1101010134 & 2018-12-18               & 58470                   & $0.04\pm0.01$ & $0.15\pm0.01$          & $135.62^{+69.44}_{-42.62}$ & $1.60\pm0.07$           & $0.24$ & 271.46/276                       \\
1101010136$^{\dagger}$ & 2019-02-18               & 58532                   & $0.04\pm0.01$ & $0.87\pm0.01$          & $7.62\pm0.21$   & $2.13\pm0.03$   & $4.22$ &628.19/756                       \\
2101010101 & 2019-04-18               & 58591                   & $0.04\pm0.01$ & $1.17\pm0.003$         & $5.41\pm0.06$                    & $-$ & $12.7$ & 1007.83/858                      \\
2101010102 & 2019-05-25               & 58628                   & $0.04\pm0.01$ & $1.21\pm0.003$         & $5.76\pm0.06$           &$-$& $15.3$ & 1086.51/868                      \\
2101010103 & 2019-06-21               & 58655                   & $0.04\pm0.01$ & $1.09\pm0.01$          & $5.73\pm0.26$     & $-$& $10.2$ & 426.17/472                       \\
2101010104 & 2019-06-22               & 58656                   & $0.03\pm0.01$ & $1.01\pm0.01$ & $5.74\pm0.15$          & $-$   & $10.1$ &609.10/638                       \\
2101010105 & 2019-07-11               & 58675                   & $0.04\pm0.01$ & $1.11\pm0.01$          & $5.44\pm0.11$   & $-$   & $10.3$ & 573.05/682                       \\
2101010106 & 2019-07-14               & 58678                   & $0.04\pm0.01$ & $1.12\pm0.01$           & $5.36\pm0.12$    & $-$  & $10.6$ & 531.87/646                       \\
2101010107 & 2019-07-15               & 58679                   & $0.04\pm0.01$ & $1.11\pm0.01$           & $5.55\pm0.15$   & $-$ & $10.7$ & 455.09/591                       \\
2101010109 & 2019-08-15               & 58710                   & $0.04\pm0.01$ & $1.11\pm0.01$ & $5.71^{+0.18}_{-0.17}$   & $-$ & $11.1$ & 415.70/543                       \\
2101010111 & 2019-11-18               & 58805                   & $0.04\pm0.01$ & $0.99\pm0.01$          & $5.98\pm0.18$    &$-$& $7.19$ & 786.82/572                       \\
2101010112 & 2019-12-04               & 58821                   & $0.04\pm0.01$ & $0.51\pm0.01$           & $-$        & $3.53\pm0.13$ &  $2.85$ & 318.86/414                       \\
2101010113 & 2019-12-05               & 58822                   & $0.04\pm0.01$ & $0.54\pm0.02$          & $   16.58^{+1.11}_{-1.38}$   & $2.30^{+0.38}_{-0.37}$  & $2.88$ & 334.59/412                    \\
2101010116 & 2019-12-13               & 58830                   & $0.07\pm0.01$ & $0.39\pm0.01$          & $23.83^{+2.32}_{-1.97}$   & $2.50^{+0.11}_{-0.12}$          & $1.28$ &294.75/388                       \\
2101010117 & 2019-12-14               & 58831                   & $0.04\pm0.01$ & $0.32\pm0.01$ & $-$        & $1.14\pm0.09$  & $1.29$  & 235.03/306                       \\
2101010120 & 2020-01-30               & 58878                   & $0.04\pm0.01$ & $0.99\pm0.01$          & $6.03\pm0.28$  & $-$ & $7.39$ & 355.65/419                       \\
2101010122 & 2020-02-18               & 58897                   & $0.03\pm0.01$ & $1.03\pm0.01$          & $5.55\pm0.11$             &$-$ & $7.84$ & 649.33/672                       \\
3609010101 & 2020-03-12               & 58920                   & $0.04\pm0.01$ & $1.16\pm0.01$          & $5.59\pm0.10$   & $-$   & $7.84$ & 685.13/732                        \\
3101010105 & 2020-05-19               & 58988                   & $0.04\pm0.01$ & $1.04\pm0.01$          & $5.96\pm0.23$    &$-$  & $8.90$ &415.91/472                       \\
3101010106 & 2020-05-20 & 58989 & $0.03\pm0.01$ & $1.06\pm0.01$ & $5.69^{+0.18}_{-0.17}$  &$-$  & $8.96$& 470.20/536 \\
3101010107 & 2020-05-21 & 58990 & $0.04\pm0.01$ & $1.05\pm0.01$ & $5.90\pm0.20$   & $-$ & $9.17$ &  413.11/522 \\
3101010108 & 2020-05-22 & 58991 & $0.04\pm0.01$ & $1.07\pm0.01$ & $5.71\pm0.19$  &$-$ & $9.51$ & 455.53/516 \\
3101010109 & 2020-05-24 & 58993 & $0.04\pm0.01$ & $1.07\pm0.01$ & $5.89\pm0.21$  &  $-$ & $9.82$ & 435.92/520 \\
3101010110 & 2020-05-25 & 58994 & $0.04\pm0.01$ & $1.05\pm0.01$ & $6.42^{+0.35}_{-0.33}$  & $-$  & $9.94$ &288.46/397 \\
3101010112 & 2020-05-27 & 58996 & $0.04\pm0.01$ & $1.09\pm0.01$ & $5.86\pm0.28$   & $-$ & $10.3$ & 364.87/442 \\
3101010113 & 2020-05-28 & 58997 & $0.04\pm0.01$ & $1.11\pm0.01$ & $5.48\pm0.14$   &$-$ & $10.4$ & 431.33/602 \\
3609010501 & 2020-07-14 & 59044 & $0.04\pm0.01$ & $1.17\pm0.01$ & $5.22\pm0.07$   &$-$ & $1.24$ & 618.01/781 \\
3609010801 & 2020-10-14 & 59136 & $0.03\pm0.01$ & $0.90\pm0.01$ & $6.11\pm0.11$  &$-$ & $5.12$ & 685.29/681 	\\ \hline
\end{tabular}
    
    \footnotesize
    { $^{\dagger}$ Fitting carried out in $0.3-8.0$ keV}
    
\end{table*}

\begin{table*}
\caption{\textit{NuSTAR} observations of LMC X-1 and LMC X-3 considered in this study. The best-fit spectral parameters are obtained by fitting Model-1 (\textit{Tbabs(ezdiskbb+laor+nthcomp)}) to the \textit{NuSTAR} energy spectra in $3-40$ keV energy band. Errors for all these estimated values are calculated with $90\%$ confidence. }
\label{tab2}  
\centering
\begin{tabular}{ccccccccc}
\hline
ObsID      & \multicolumn{1}{c}{Date} & \multicolumn{1}{c}{MJD} & \multicolumn{1}{c}{$T_{max}$}  & \multicolumn{1}{c}{$N_{ezd}$}     & $\Gamma$                          & \multicolumn{1}{c}{lineE}    & $F_{bol}$ \footnotesize{$\times10^{-10}$}              & $\chi^{2}/dof$                  \\
 & 
& & \multicolumn{1}{c}{\footnotesize{(keV)}}& \multicolumn{1}{c}{} &  & \multicolumn{1}{c}{\footnotesize{(keV)}} & (\footnotesize{erg cm}$^{-2}$ s$^{-1}$) & \\ \hline
\multicolumn{8}{c}{LMC X-1}                                                                                                                    \\ \hline
30001039002 & 2014-08-08 & 56877  & $0.84\pm0.01$ & $6.98^{+0.44}_{-0.40}$ & $3.53^{+0.18}_{-0.17}$ &  $7.11^{+0.10}_{-0.13}$ & $2.89$ & $302.66/560$  \\
30001143002 & 2014-11-09 & 56970  & $0.87\pm0.01$ & $9.42^{+0.34}_{-0.36}$ & $2.92\pm0.04$ &  $7.18^{+0.09}_{-0.07}$ & $5.04$ & $1262.18/923$  \\
30201029002 & 2016-05-12 & 57520  & $0.80\pm0.01$  & $14.34^{+0.64}_{-0.59}$ & $2.68\pm0.05$    &    $7.16\pm0.09$   & $4.69$ & $1161.48/849$     \\
\hline
\multicolumn{8}{c}{LMC X-3}                                                                                                                    \\ \hline
30101052002 & 2015-12-29 & 57385 & $1.10\pm0.01$ & $5.25\pm0.09$ & $2.27\pm0.11$  & $-$ & $6.19$ & $534.75/559$ \\
30402035002 & 2019-02-18 & 58532            & $1.00\pm0.01$  & $4.44\pm0.16$          & $2.35^{+0.05}_{-0.14}$     &  $-$     & $4.35$ & $702.24/608$   \\
10601308002 & 2020-04-07 & 58946           & $0.88\pm0.01$ & $3.62^{+0.33}_{-0.29}$ & $2.12^{+0.14}_{-0.13}$  &$-$ & $1.63$ & $389.15/345$   \\ \hline      
\end{tabular}
\end{table*}

\subsection{Estimation of Physical Parameters}
\label{sec4.2}
Based on the results of spectral analysis presented in section \ref{sec4.1.1} and \ref{sec4.1.2}, it is understood that during most of the \textit{NICER} and \textit{NuSTAR} observations considered, both sources are in soft state and hence we attempt to constrain the physical parameters of these sources by using continuum-fitting method as discussed in section \ref{sec3.2}.
By modelling the \textit{NICER} spectral data using Model-2, the source spin value is estimated to be $0.90\pm0.01$ for LMC X-1 and within $0.23-0.33$ for LMC X-3 from this continuum-fitting method. The estimated $\dot{M}$ of LMC X-1 during these fits is $1.79^{+0.05}_{-0.07} \times 10^{18}$ g s$^{-1}$ and in case of LMC X-3, $\dot{M}$ during different observations varies in the range $4.28-5.57 \times10^{18}$ g s$^{-1}$. Similar fitting for the \textit{NuSTAR} spectra resulted in $a=0.93\pm0.01$ for LMC X-1 and $0.16-0.28$ for LMC X-3. These fits resulted in $\dot{M}$ of $0.90-1.83 \times 10^{18}$ g s$^{-1}$ for LMC X-1 and $1.82-6.56 \times 10^{18}$ g s$^{-1}$ for LMC X-3. Parameters obtained from Model-2 fit along with its error for different observations are listed in Table \ref{tab4}.  In Figure \ref{fig8}, we show the statistical variation of $\Delta \chi^{2}$ with respect to the different values of $a$ for both sources obtained from Model-2 fitting to the \textit{NuSTAR} energy spectra. $\Delta\chi^{2}$ values are obtained using \texttt{steppar} command in \texttt{XSPEC} after loading the MCMC chain. The horizontal line plotted represents the 90$\%$ confidence range (i.e. $\Delta \chi^2 = 2.706$). \par
\begin{figure}
    \centering
    \includegraphics[width=9cm]{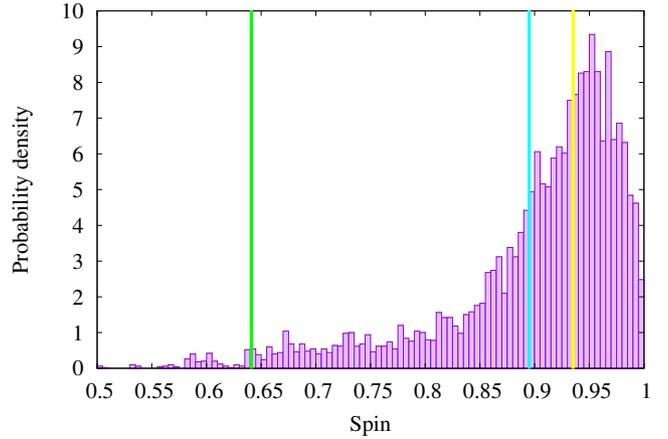}
    \caption{Probability distribution of spin parameter obtained from \textit{relline} model in Model-3 fitting to the \textit{NuSTAR} spectrum of LMC X-1. Probability density is obtained using MCMC chain results in \texttt{XSPEC}. The plotted data belongs to observation on MJD 56877. Error confidence of $1\sigma$ (68.3$\%$), $2\sigma$ (95.4$\%$) and 3$\sigma$ (99.7$\%$) are marked in green, cyan and yellow colors respectively.}
    \label{fig9}
\end{figure}

From Model-3 fitting to the \textit{NuSTAR} spectra, we could constrain the $R_{in}$ to be $\sim1.35\pm0.70$ R$_{g}$ when $a$ and $a_{*}$ are mutually tied. Model-3 fitting carried out by freezing the $R_{in}=1.35$ R$_{g}$ and two spin parameters to vary independently results in $a=0.85-0.94$ and $a_{*}=0.93-0.94$. Parameter values from these fits are listed in Table \ref{tab5}. Probability distribution of $a_{*}$ obtained from Model-3 fitting to the MJD 56877 spectrum is shown in Figure \ref{fig9}. Different confidence levels i.e., $1\sigma$ (68.9$\%$), $2\sigma$ (95.4$\%$) and $3\sigma$ (99.7$\%$) are marked for the final estimation of the spin from the fit. From the values listed in Table \ref{tab5}, it can be seen that spin value obtained from Fe-line fitting method is within the error range of value obtained from continuum-fitting method.   \par

\begin{table*}
\caption{\textit{AstroSat} observations of LMC X-1 and LMC X-3 considered in this study. The best-fit spectral parameters are obtained by fitting Model-1 (\textit{Tbabs(ezdiskbb+nthcomp)}) to the \textit{AstroSat} energy spectra in 0.5$-$20 keV energy range. Errors for all these estimated values are calculated with $90\%$ confidence.}
\centering
\label{tab3}
\begin{tabular}{ccccccccc}
\hline
ObsID                     & Date         & MJD   & $n_{H}$                & $T_{max}$     & $N_{ezd}$              & $\Gamma$               & $F_{bol}$ \footnotesize{$\times10^{-10}$}                 & $\chi^{2}/dof$ \\
& & & \footnotesize{$\times10^{22}$ cm$^{-2}$} & \footnotesize{(keV)} & & & \footnotesize{erg cm$^{-2}$ s$^{-1}$}  & \\

\hline
\multicolumn{9}{c}{LMC X$-$1}                          \\ \hline
0826 & 25-11-2016 & 57717 & $1.07^{+0.03}_{-0.02}$ & $0.83\pm0.01$ & $8.45^{+1.08}_{-1.24}$ & $4.08^{+0.11}_{-0.26}$ & $3.38$ & $700.35/529$   \\
1496 & 28-08-2017 & 57993 & $1.05^{+0.08}_{-0.05}$ & $0.84\pm0.02$ & $11.05\pm1.33$         & $2.66\pm0.12$          & $3.92$ & $606.19/417$   \\
3414 & 01-06-2020 & 58854 & $1.14^{+0.02}_{-0.03}$ & $0.90\pm0.01$ & $5.89^{+0.84}_{-3.91}$ & $4.98^{+0.84}_{-3.73}$ & $3.40$ & $547.79/480$   \\
3596 & 04-01-2020 & 58939 & $1.12^{+0.05}_{-0.03}$ & $0.90\pm0.01$ & $8.05^{+0.15}_{-0.63}$ & $2.53^{+0.80}_{-0.39}$ & $2.95$ & $675.26/506$   \\
3596 & 04-01-2020 & 58940 & $1.2^{+0.12}_{-0.11}$ & $0.85\pm0.03$ & $8.05^{+1.49}_{-0.70}$ & $3.50\pm0.36$          & $3.68$ & $384.41/338$   \\ \hline
\multicolumn{9}{c}{LMC X$-$3}                \\ \hline
0600 & 14-08-2016 & 57614 & $0.03\pm0.01$          & $1.03\pm0.01$ & $3.45^{+0.18}_{-0.19}$ & $-$                    & $2.30$ & $473.29/403$   \\
0672 & 19-09-2016 & 57650 & $0.02\pm0.01$          & $1.03\pm0.01$ & $3.71^{+0.36}_{-0.30}$ & $-$                    & $2.41$ & $455.65/387$   \\
1130 & 03-04-2017 & 57846 & $0.03\pm0.01$          & $0.94\pm0.01$ & $4.55^{+0.26}_{-0.10}$ & $-$                    & $1.93$ & $446.33/417$   \\
1190 & 19-04-2017 & 57862 & $0.02^{+0.02}_{-0.01}$ & $0.98\pm0.01$ & $3.35^{+0.24}_{-0.21}$ & $-$                    & $1.78$ & $338.24/318$   \\
1698 & 17-11-2017 & 58074 & $0.04\pm0.01$          & $1.04\pm0.01$ & $3.86^{+0.11}_{-0.08}$ & $-$                    & $5.34$ & $525.21/425$   \\
1778 & 17-12-2017 & 58104 & $0.04\pm0.01$          & $1.07\pm0.01$ & $4.23^{+0.53}_{-0.18}$ & $-$                    & $3.35$ & $450.16/452$   \\
1884 & 06-02-2018 & 58155 & $0.05\pm0.01$          & $1.08\pm0.01$ & $4.26^{+0.18}_{-0.26}$ & $-$                    & $3.83$ & $597.38/453$   \\
1908 & 20-02-2018 & 58169 & $0.04\pm0.01$          & $1.08\pm0.01$ & $4.48^{+0.48}_{-0.26}$ & $-$                    & $3.72$ & $532.75/447$   \\
3430 & 12-01-2020 & 58860 & $0.02\pm0.01$          & $0.98\pm0.01$ & $3.56^{+0.05}_{-0.09}$ & $-$                    & $1.89$ & $669.34/480$   \\ \hline
\end{tabular}
\end{table*}
\begin{table*}
\caption{Best-fit parameters obtained by fitting Model-2 (\textit{Tbabs(simpl*kerrbb+laor)}) to the \textit{NICER} and \textit{NuSTAR} energy spectra of LMC X-1 and LMC X-3. Here $\Gamma$ and $FracSctr$ represents the photon index and fraction of scattered photons that is obtained from \textit{simpl} model. $a$ and $\dot{M}$ represents spin and mass accretion rate parameters. $lineE$ is the line energy obtained from the \textit{laor} model. For these fits, distance to the source is frozen to $48.1$ kpc for both sources. Inclination angle of LMC X-1 and LMC X-3 are chosen as $36.4^{\circ}$ and $69.4^{\circ}$ respectively. $M_{BH}$ is fixed to $8.82 M_{\odot}$ for LMC X-1 and $6.48 M_{\odot}$ for LMC X-3. Errors of $a$ and $\dot{M}$ are calculated using MCMC simulation method.}
\centering
\label{tab4}
% \usepackage{multirow}
% \usepackage{multirow}
\begin{tabular}{cccccccc}
\hline
                        & ObsID       & $\Gamma$               & $FracSctr$                   & $a$           & $\dot{M} \times10^{18}(g s^{-1})$ & $lineE$ (keV) & $\chi^{2}/dof$ \\ \hline
\multirow{7}{*}{NICER}  & \multicolumn{6}{c}{LMC X-1}                                                                                                         \\ \cline{2-8} 
                        & 1100070101  & -                      & -                            & $0.90\pm0.01$ & $1.79^{+0.05}_{-0.07}$  & $-$     & 627.40/589     \\ \cline{2-8} 
                        & \multicolumn{6}{c}{LMC X-3}                                                                                                         \\ \cline{2-8} 
                        & 1101010120  & -                      & -                            & $0.23\pm0.02$ & $4.53^{+0.07}_{-0.08}$             & $-$  & 521.27/545     \\
                        & 1101010121  & -                      & -                            & $0.23\pm0.02$ & $4.28\pm0.07$              & $-$  & 389.96/503     \\
                        & 2101010105  & -                      & -                            & $0.33\pm0.01$ & $5.57\pm0.06$              & $-$  & 599.67/679     \\
                        & 3101010108  & -                      & -                            & $0.33\pm0.02$ & $5.11\pm0.09$              & $-$  & 443.28/513     \\ \hline
                        & \multicolumn{6}{c}{LMC X-1}                                                                                                         \\ \cline{2-8} 
\multirow{7}{*}{NuSTAR} & 30001039002 & $3.47^{+0.19}_{-0.13}$          & $0.09\pm0.02$ & $0.94^{+0.01}_{-0.02}$ & $0.90\pm0.05$                & $7.01^{+0.17}_{-0.16}$ & $314.91/621$   \\ 
                        & 30001143002 & $2.92\pm0.05$          & $0.13\pm0.01$ & $0.88\pm0.01$ & $1.58\pm0.05$   & $7.24\pm0.09$             & $998.81/793$   \\ 
                        & 30201029002 & $2.68^{+0.08}_{-0.04}$          & $0.10\pm0.01$ & $0.85\pm0.02$ & $1.83\pm0.09$  & $7.11\pm0.10$              & $1063.01/842$   \\ \cline{2-8} 
                        & \multicolumn{6}{c}{LMC X-3}                                                                                                         \\ \cline{2-8} 
                        & 30101052002 & $1.73^{+0.14}_{-0.13}$ & $0.01\pm0.005$               & $0.16\pm0.01$ & $6.56^{+0.10}_{-0.09}$ & $-$      & $630.52/559$   \\
                        & 30402035002 & $2.29\pm0.07$          & $0.08\pm0.005$               & $0.18\pm0.03$ & $4.31^{+0.14}_{-0.11}$  & $-$     & $682.17/603$   \\
                        & 10601308002 & $2.00\pm0.15$          & $0.05\pm0.005$               & $0.28\pm0.05$ & $1.82\pm0.13$      &   $-$       & $380.61/355$   \\ \hline
\end{tabular}
\end{table*}
% \usepackage{multirow}
\begin{table*}
\caption{Spectral parameters obtained by fitting Model-3 (\textit{Tbabs(simpl*kerrbb+relline)}) to the \textit{NuSTAR} spectra of LMC X-1. Here, $a$ represents the spin parameter obtained from \textit{kerrbb} and $a_{*}$ is the spin from \textit{relline}. Errors are estimated with $90\%$ confidence using MCMC simulation method. }
\centering
\label{tab5}
\begin{tabular}{lllll}
\hline
Model                     & Parameter       & MJD 56877               & MJD 56970              & MJD 57520              \\ \hline
\multirow{2}{*}{\texttt{kerrbb}} & $a$  & $0.94\pm0.01$           & $0.93\pm0.01$          & $0.85^{+0.02}_{-0.01}$          \\
                          & $\dot{M}\times10^{18}$ (g/s)           & $0.90^{+0.06}_{-0.05}$ & $1.42^{+0.05}_{-0.04}$ & $1.83^{+0.10}_{-0.06}$ \\
\texttt{relline}                   & lineE (keV)     & $7.10^{+0.09}_{-0.16}$  & $7.27^{+0.11}_{-0.10}$ & $7.21^{+0.07}_{-0.11}$ \\
                          & $a_{*}$               & $0.94^{+0.04}_{-0.24}$  & $0.93^{+0.04}_{-0.12}$ & $0.93^{+0.04}_{-0.06}$ \\
\texttt{simpl}                   & $\Gamma$        & $3.46^{+0.19}_{-0.13}$           & $2.92^{+0.18}_{-0.14}$ & $2.68^{+0.08}_{-0.04}$ \\
                          & $FracSct$   & $0.09\pm0.02$                   & $0.13\pm0.01$                  & $0.10\pm0.01$ \\ \hline
                          
                          & $\chi^{2}/dof$ & $314.11/621$ & $989.39/785$ & $1053.41/841$ \\
                          \hline
\end{tabular}
\end{table*}
\subsection{Temporal Properties}
\label{sec4.3}
As explained in section \ref{sec3.3}, we estimated $F_{var}$ of LMC X-1 and LMC X-3 using long-term monitoring observations by \textit{MAXI}. $F_{var}$ of $\sim 20\%$ is obtained for LMC X-1 whereas LMC X-3 shows a relatively high fluctuation of source flux with $F_{var}\sim49.74\%$. Lightcurves of LMC X-1 and LMC X-3 plotted using \textit{MAXI} observational data (Figure \ref{fig2}) for the period of $2014-2020$ illustrates this moderate variability of LMC X-1 and high variability of LMC X-3 in long-term scale. In order to understand the short-term variability, we examine the lightcurves belonging to different \textit{NICER} and \textit{NuSTAR} observations. \textit{NICER} and \textit{NuSTAR} PDS of LMC X-1 and LMC X-3 are fitted using the \textit{powerlaw} model. In order to quantify the variability, fractional rms is estimated from the PDS as explained in section \ref{sec3.3}. Fractional rms of LMC X-1 obtained from \textit{NICER} PDS in the energy range $0.3-10$ keV and frequency range $0.002-10$ Hz during the two observations are $1.81\%$ and $2.26\%$. In case of LMC X-3, during soft states, rms in $0.3-10$ keV band is found to be $\sim0.08-2.35\%$. During the intermediate states, rms increases to $3.05-4.91\%$ and a high variability of $17.06\%$ is obtained during hard states of LMC X-3. However, \textit{NuSTAR} PDS is found to have power only till 1 Hz and hence rms is not estimated for the same. We also examine the PDS to check for the presence of QPOs in low frequency as well as in high frequency range, but do not find any signature during the different spectral states. 

\section{Discussion and Conclusion}
\label{sec5}
The extragalactic BH-XRBs LMC X-1 and LMC X-3 are widely studied using observational data from different instruments since its detection. While these studies revealed a lot of information about these sources, availability of the observations using existing observatories which consists of various instruments with better effective area, wide-band coverage and time resolution provides a scope to explore the source characteristics in detail. Therefore in this work, we make use of multi-epoch X-ray observations from various current observatories and perform `spectro-temporal' analysis of extragalactic BH-XRBs LMC X-1 and LMC X-3. The observations performed by \textit{NuSTAR}, \textit{NICER} and \textit{AstroSat} during their era which spans over the period of 2014$-$2020 are considered for this study. \par
It is evident from section \ref{sec4.1.1}, \ref{sec4.1.2}, Tables \ref{tab1}, \ref{tab2} and \ref{tab3} that, LMC X-1 exhibits a steady spectral state during this entire period whereas LMC X-3 undergoes state change. Spectral analysis of LMC X-1 shows that the source energy spectra are characterized by three spectral components: thermal disc blackbody, Comptonized emission and Fe K$\alpha$ line emission. Steep spectral index ($\Gamma\sim3$) along with higher percentage of disc flux contribution ($>79\%$) seen by \textit{NuSTAR}, \textit{NICER} and \textit{AstroSat} observations during 2014$-$2020 suggest that the source was in a steady soft state for this entire period. Source spectrum and its flux remain almost constant (see Figure \ref{fig4}) throughout the period with no change in the spectral state. Temporal properties agree with the soft nature of the source by having very low rms of $\sim0.08-2.35\%$ in $0.3-10$ keV. This behaviour is similar to the results obtained from previous X-ray missions such as \textit{BeppoSAX}, \textit{RXTE} and \textit{Suzaku} where the source is always found in its 
soft state \citep{2001ApJS..133..187H,2001MNRAS.320..316N,2015PASJ...67...46K}. \par
In case of LMC X-3, the source
is found to have undergone soft-to-intermediate-to-hard state transition quite a few times within the period of $2015-2020$. Spectral analysis show that the source spectra are well described by a thermal disc blackbody along with a Comptonization component (see Figure \ref{fig6} and \ref{fig7}). While Comptonized component is present along with thermal component in all the \textit{NuSTAR} spectra, it is present only during intermediate and hard state observations in the \textit{NICER} observations i.e. when $F_{bol}$ is $<4\times10^{-10}$ erg cm$^{-2}$ s$^{-1}$ and $f_{disc}<75\%$. Accretion disc is found to have almost constant high temperature during all the soft state observations ($T_{max}\sim1$ keV). As the spectrum becomes harder, disc is seen to have cooled down i.e. $T_{max}=0.38-0.87$ keV and $\sim0.16$ keV during intermediate hard states respectively. This change in temperature during hard state observation is consistent with the results obtained by \cite{2001MNRAS.320..327W} using \textit{RXTE} observations of LMC X-3. Disc normalization, which is related to inner disc radius $R_{in}$ is found to be nearly constant while in the soft state, however it increases during the intermediate and hard states (see Table \ref{tab1} and \ref{tab2}). Normalization remaining constant in the high state is consistent with previous studies on LMC X-3 \citep{2001MNRAS.320..327W, 2010ApJ...718L.117S} as well as with Galactic BHs such as Cyg X-1 and GX 339-4 \citep{2011ApJ...742...85G,gx339megumi}. This has been interpreted as $R_{in}$ closely related to $R_{ISCO}$. Spectral index of the Comptonization component as seen by \textit{NICER} spectral fits has a value of $\Gamma=2.02-2.36$ during intermediate state and it decreases to $\sim1.6$ during the hard state. This is consistent with the generally found trend in BH-XRBs i.e. softer the spectrum, steeper is the spectral index. Correlation is seen between the source flux, thereby mass accretion rate and the spectral state. As explained in Section \ref{sec4.1.2}, source flux in $0.3-40$ keV is higher during soft state ($\sim1.35\times10^{-9}$ erg cm$^{-2}$ s$^{-1}$), while it decreased during intermediate state ($\sim2.7\times10^{-10}$ erg cm$^{-2}$ sec$^{-1}$) and further in hard state ($\sim0.24\times10^{-11}$ erg cm$^{-2}$ sec$^{-1}$) (see Figure \ref{fig7}). Mass accretion rate calculated using bolometric luminosity in $0.1-50$ keV from \textit{NICER} is high during soft state ($0.16-0.42$ $\dot{M}_{Edd}$) and it gradually decreases during intermediate ($\sim0.09$ $\dot{M}_{Edd}$) and it reaches minimum during hard state ($<0.01$ $\dot{M}_{Edd}$). Such a correlation is consistent with the model proposed by \cite{1997ApJ...489..865E} where change in accretion rate is explained to be the driving mechanism for the spectral state change. Correlation is also seen between the contribution of disc flux and source flux. Hence, we classify the spectra with $f_{disc}>75\%$ as soft state, $f_{disc}=35-75\%$ as intermediate state and $f_{disc}<35\%$ as hard state. Clear correlation between the disc flux and total source flux, indicates that the disc flux drives the spectral variability in LMC X-3.  
Temporal analysis resulted in a very low rms of $\sim0.08-2.35\%$ for PDS obtained in $0.3-10$ keV. The rms increased moderately to $3.05-4.91\%$ in the intermediate state, while in hard state relatively high rms of $17.06\%$ is obtained. These temporal properties further support our classification of spectral states. This result is consistent with the understanding that BH binaries show very minimal variability during high soft state in low energy \citep{2005A&A...440..207B,nandi2012accretion}. \par
From the collective analysis of \textit{NICER}, \textit{NuSTAR} and \textit{AstroSat} observations of LMC X-3, it can be seen that within the period of 2015$-$2020, LMC X-3 has undergone transition into hard state once during MJD 58470 and to intermediate state four times i.e during MJD 58463, 58832, 58821$-$58831 and 58946. During rest of the observations source remained consistently in hard state. Among these, observations done on MJDs 58821 and  58831 have energy spectra which are well described by the Comptonization model alone. This could be due to the low statistic which does not allow us to detect the disc component. However these cannot be classified into hard state because of steep spectral index (see Table \ref{tab1}) and the relatively high flux. Since its flux is close to that of intermediate observations (see Section \ref{sec4.1.1}) we consider these two observations as belonging to intermediate state. 

\par
\cite{2001MNRAS.320..327W} have carried out similar spectral transition study of LMC X-3 using \textit{RXTE} observations and they found a correlation between \textit{ASM} count rate and the state transition. According to their results, when the \textit{ASM} count rate decreases $<6$ counts sec$^{-1}$, source transits from its typical soft state to hard state. Similar behaviour is seen in our study as well where we observe that during the epoch when source transited into hard state, there is a decrease in \textit{MAXI} count rate, while it remains high during the soft state (see Figure \ref{fig2}). This indicates that the long term spectral variability of LMC X-3 (see Section \ref{sec4.3}) is correlated to the change in spectral state. This behaviour is also seen in Galactic BH-XRBs such as Cyg X-1 which shows spectral state is correlated with evolution of lightcurve \citep{2006A&A...446..591C}. Most widely accepted explanation for change in spectral variability of persistent sources in Galactic BHs is the change in accretion disc due to mass accretion \citep{1998ApJ...505..854E}. Change in luminosity as a result of variation in accretion rate during different spectral states is validated by our spectral analysis (see Section \ref{sec4.1.1} and \ref{sec4.1.2}).
\par
Stable luminosity ($\sim1.0\times10^{-9}$ erg cm$^{-2}$ sec$^{-1}$) in low energy range in the soft state \textit{NICER} spectra of LMC X-1 allowed us to constrain the physical parameters by fitting the relativistic accretion disc model. This resulted in $a=0.90\pm0.01$ (Table \ref{tab4}) from \textit{NICER} fits. Similar analysis of the \textit{NuSTAR} spectra resulted in $a=0.85-0.94$ (Table \ref{tab4}, Figure \ref{fig8}). This high spin value is consistent with that reported by \cite{2009ApJ...701.1076G} using \textit{RXTE} observations, \cite{2020MNRAS.498.4404M} and \cite{2021MNRAS.501.5457B} using \textit{AstroSat} observations. \cite{2021arXiv210810546J} has recently reported the spin of LMC X-1 to be within $0.92-0.95$ using \textit{Swift} and \textit{NuSTAR} observations. Thus, the spin value of LMC X-1 we obtained is consistent across the results presented till date over the X-ray energy range. We also estimate the spin of LMC X-1 using Fe-line fitting method from \textit{NuSTAR} energy spectra which is found to be $0.93-0.94$ (Table \ref{tab5}, Figure \ref{fig9}). This shows the consistency between continuum-fitting and Fe-line fitting method of spin estimation. In addition, this value is also consistent with the previous spin estimation using Fe-line fitting method by \citep{2012MNRAS.427.2552S}. 
\par
Similar continuum-fitting analysis of soft state spectra of LMC X-3 resulted in $a=0.23-0.33$ from \textit{NICER} and $0.16-0.28$ from \textit{NuSTAR} (see Table \ref{tab4}, Figure \ref{fig8}). This is consistent with the value estimated using \textit{RXTE} observations by \cite{2014ApJ...793L..29S}, \textit{AstroSat} observations in \cite{2021MNRAS.501.5457B} and \textit{Swift+NuSTAR} observations by \cite{2021arXiv210810546J}. Consistency across continuum-fitting and Fe-line fitting method could not be checked in LMC X-3 because of absence of Fe emission line in its spectra.
\par
As explained in Section \ref{sec4.2}, 
attempt to constrain the high energy reflection parameter in the \textit{NuSTAR} and \textit{AstroSat} energy spectra using relativistic reflection model was not fruitful. Even though \textit{relxill} model resulted in fit with acceptable $\chi^{2}_{dof}$ for both sources, we could not constrain any of the physical parameters such as spin, reflection fraction and ionization parameters from this fit. \cite{2021arXiv210810546J} has reported a weak reflection in LMC X-1 and absence of reflection in LMC X-3 using \textit{Swift} and \textit{NuSTAR} data which we could not confirm in our study due to the inability to constrain \textit{relxill} model parameters.

To summarize the present work:
\begin{itemize}
    \item{Long term variability and HR studied from \textit{MAXI} lightcurve show that over the years, LMC X-1 has a moderate flux variability ($F_{var}\sim20\%$) whereas LMC X-3 shows higher variability ($F_{var}\sim50\%$).}
    \item{LMC X-1 is found to be in a steady soft state for the whole period of $2014-2020$, whereas a state transition is seen in LMC X-3 in the following sequence: \textit{soft $\rightarrow$ intermediate $\rightarrow$ hard $\rightarrow$ intermediate $\rightarrow$ soft $\rightarrow$ intermediate $\rightarrow$ soft state} during this period.}
    \item{Spectral properties of sources during the different spectral states are similar to that of a typical BH with thermal disc dominant spectra in soft state, Comptonized flux dominant spectra in hard state and an intermediate state whose properties falls in between hard and soft state.}
    \item{Temporal properties indicate constant low value of fractional rms in PDS of LMC X-1 whereas the rms increases in LMC X-3 as it transits from soft state towards a harder state.}
    \item{BH spin estimation using  continuum-fitting yields $a=0.85-0.94$ for LMC X-1 and $0.16-0.33$ for LMC X-3. Spin of LMC X-1 estimated from Fe-line fitting method is $0.93-0.94$. This shows the consistency of spin value across observations from different X-ray instruments as well as across two different methods of spin estimation.}
\end{itemize}

\section{Acknowledgments}
We thank the reviewers for their
feedback and comments which helped to improve the quality of this paper. The authors acknowledge the financial support of Indian Space Research Organization (ISRO) under \textit{AstroSat} archival data
utilization program Sanction order No. DS-2B-13013(2)/13/2019-Sec.2. This  work  made use of data from the \textit{NuSTAR} and \textit{NICER} mission by the National Aeronautics and Space Administration. This research has made use of the \textit{NuSTAR} Data Analysis Software (NuSTARDAS) and NICER Data Analysis Software (NICERDAS). This publication uses data from the {\it AstroSat} mission of the ISRO archived at the Indian Space Science Data Centre (ISSDC).
This work has been performed utilizing the calibration databases and
auxiliary analysis tools developed, maintained and distributed by
{\it AstroSat-SXT} team with members from various institutions in India
and abroad. This research has made use of MAXI data provided by RIKEN, JAXA and the MAXI team. Also this research made use of software provided by the High Energy Astrophysics Science Archive Research Center (HEASARC) and NASA’s Astrophysics Data System Bibliographic Services. AN thanks GH, SAG; DD, PDMSA and Director, URSC for encouragement and continuous support to carry out this research.

%% Bibliography
%% Author year style
\bibliographystyle{model5-names}
\biboptions{authoryear}
\bibliography{refs.bib}
\end{document}